\documentclass[aip,reprint,amsmath,amssymb]{revtex4-2}
\usepackage{graphicx}
\usepackage{dcolumn}
\usepackage{bm}
\usepackage[utf8]{inputenc}
\usepackage[T1]{fontenc}
\usepackage{mathptmx}
\usepackage{etoolbox}
\usepackage{mathtools}
\usepackage[colorlinks,allcolors=blue]{hyperref}
\usepackage{gensymb}
\usepackage{amsmath,txfonts}
\usepackage{esint}
\usepackage[export]{adjustbox}
\usepackage{siunitx}
\usepackage{float}
\usepackage{changepage}

\DeclareMathAlphabet{\mathcal}{OMS}{cmsy}{m}{n} 
\DeclareUnicodeCharacter{2212}{-}
\makeatletter
\def\@email#1#2{%
	\endgroup
	\patchcmd{\titleblock@produce}
	{\frontmatter@RRAPformat}
	{\frontmatter@RRAPformat{\produce@RRAP{*#1\href{mailto:#2}{#2}}}\frontmatter@RRAPformat}
	{}{}
}%
\makeatother
\begin{document}
	
	\preprint{AIP/123-QED}
	
	\title[]{ Effect of Magnetic Field on Aqueous Humor Flows Inside Anterior Chamber of Human Eye}
	\author{Deepak Kumar}
	\author{Subramaniam Pushpavanam $^*$}%
	\email{spush@iitm.ac.in}
	\affiliation{ Department of Chemical Engineering, Indian Institute of Technology Madras, Chennai 600036, India}
	%

	\date{\today }
	
	\begin{abstract}
		Aqueous humor (AH) dynamics is responsible for maintaining intraocular pressure, ocular health and targeted drug delivery within the eye. This study investigates the flow of AH within the anterior chamber (AC) under the combined influence of a uniform magnetic field and natural convection. Different orientations of the magnetic field and temperature gradient are considered. A lubrication approximation is employed and the resulting equations are solved using regular perturbation method. The analytical solutions are validated using numerical simulations performed in COMSOL Multiphysics 6.2\textsuperscript \textregistered. In the standing position, AH flow field is characterised by a single vortex, while in the supine position, it forms two counter-rotating vortices. The velocity is found to be higher in standing position. The effect of a uniform magnetic field on the velocity is more significant in the supine position. The magnetic field does not change the flow field qualitatively as buoyancy is the primary driving force. In the standing position a magnetic field oriented perpendicular to the eye resulted in a greatest reduction of AH velocity, as compared to a magnetic field along the eye. This study is a step towards holistic approach for targeted drug delivery using magnetic fields in eye. 
	\end{abstract}
	\maketitle
	\section{\label{secI}Introduction}
	The circulation of aqueous humor (AH) inside the anterior chamber (AC) is essential for maintaining the health and functionality of eye. It provides necessary nutrients and removes waste products from various ocular tissues \citep{sacca2014focus}.  A temperature difference across the iris and the cornea induces buoyancy driven flow. Several studies have focused on the effect of natural convection in AH flow by considering density of AH as a function of temperature. These models use the Boussinesq approximation \citep{canning2002fluid,heys2002boussinesq}. They help understand mixing of AH and formation of Krukenberg spindle on the corneal surface. Several numerical models have been developed to examine the effect of heat transfer on AH flow. The temperature distribution in human and rabbit eyes under various heating scenarios was analysed using a finite difference approach \citep{lagendijk1982mathematical}. A 2-D numerical model was used to investigate the effect of AH flow on temperature distribution inside the AC using finite element and boundary element methods \citep{ng2006fem,ooi2008simulation,ooi2011effects}. 
	Pigmentary glaucoma is characterized by accumulation of pigmented particles within the AC. These particles originate from the iris and are released into the AH. They can also cause potential blockages and alter fluid flow \citep{alvarado1992outflow,bustamante2021pigment,okafor2017update}. This is determined by the shear stresses generated by the flow. Kaji et al investigated the effect of shear stress on corneal endothelial cell loss due to changes in hydrodynamics after laser iridectomy. They observed that corneal cell detachment occurred at shear stress level exceeding $0.3$  \si{dyne/cm^2} \citep{kaji2005effect}. Shear stress on corneal endothelial were found to be greater for shallower ACs \citep{yang2024anterior}.
	
	AH is produced by ciliary body and it consists of ions such as $Na^+,H^+,K^+,HCO_3^-$ and $Cl^-$ \citep{sears2022aqueous}. Consequently, the AH is a conducting fluid \citep{vadde2023review}. AH increasingly resembles plasma as it passes from posterior to anterior chamber of eye due to exchange of nutrients \citep{yang2024anterior}. A conducting fluid in the presence of an external magnetic field experiences a Lorentz force.  This along the streamwise direction can increase friction drag and reduce the flow rate whereas Lorentz force along the spanwise direction reduces drag and thus can increase flow rate \citep{altintacs2017direct}. Magnetic fields have been used widely in microbiology, biotechnology and biomedicine in targeted therapeutic drugs \citep{berry2009progress,ziarani2019role,pankhurst2009progress,tietze2015magnetic}. They have also been used to slow down the flow of blood and supress tumour growth \citep{mckay2007literature,tatarov2011effect}. Drochon et al. reviewed the analytical solutions of magnetohydrodynamic flow of blood highlighting its potential in drug targeting, magnetic resonance imaging, blood pulse energy harvesting and tissue engineering \citep{drochon2018review}. A new methodology for targeted drug delivery in human eye using a system of magnets has been proposed to provide the maximal magnitude of the magnetic field gradient \citep{erokhin2018magnetic}. A recent experimental study focused on the effect of magnetic field in drug targeting to the eye tissues \citep{zahn2020investigation}. They found that a magnetic field gradient of 20 T/m was sufficient to allow nanoparticles to penetrate the eye tissues. 
	The aqueous humor (AH) in the anterior chamber of the eye is conductive due to the presence of ions. It has an electrical conductivity  of $1.62$ S/m \citep{lee2022vivo,lee2024multi}. The buoyancy driven AH flow can be influenced by external magnetic fields since AH is conducting. It is hence important to understand the role of magnetic fields on AH dynamics to obtain a comprehensive holistic perspective for intraocular pressure management. It would help optimize magnetic drug delivery systems and guarantee the safety of targeted ocular therapies.
	The depth of AC in human eye is significantly smaller than its width. This has been exploited and a lubrication approximation has been applied to study the flow of AH \citep{canning2002fluid, fitt2006fluid}. In these studies, they consider the pressure to be hydrostatic based on average temperature of cornea and iris. In this work, we relax this assumption and we use the fact that pressure is an intensive property to derive an additional condition on pressure. This provides a more accurate analysis of the AH flow dynamics within the AC.
	
	AH dynamics plays important role in spreading drug molecules to different ocular sites within the AC. Traditional drug delivery methods face challenges due to ocular barriers such as tear film turnover, cornea, blood aqueous barrier and blood retinal barrier \citep{bhandari2021ocular,mofidfar2021drug}. External magnetic fields provide a way to guide drug-loaded particles precisely within the AC. Therefore, understanding AH dynamics in the presence of external magnetic field is very important as it can provide an important mechanism for non-invasive targeted drug delivery. Our extensive literature survey revealed that the dynamics of AH have not been studied in the presence of an external magnetic field. This paper aims to address this gap by studying the effect of an external magnetic field on AH dynamics analytically using regular perturbation method \citep{garg2014vertically,holmes2012introduction,vinze2021effect}. Numerical simulations are also carried out using COMSOL Multiphysics 6.2\textsuperscript \textregistered to analyse realistic eye geometries.
	
	The paper is organized in the following manner. Section \ref{secII} describes the model and governing equations. Here, we justify lubrication approximation for human eye. In section \ref{secIII}, we derive the asymptotic solution using the regular perturbation method for supine and standing eye position. Different magnetic field orientations are considered. We study the effect of magnetic field on AH dynamics, with potential application in mind. Finally, in Section \ref{secIV}, we summarise the main findings and contributions of our work.
	\section{\label{secII}Model description and governing equations}
	\begin{figure*}
		\includegraphics[width=1\textwidth]{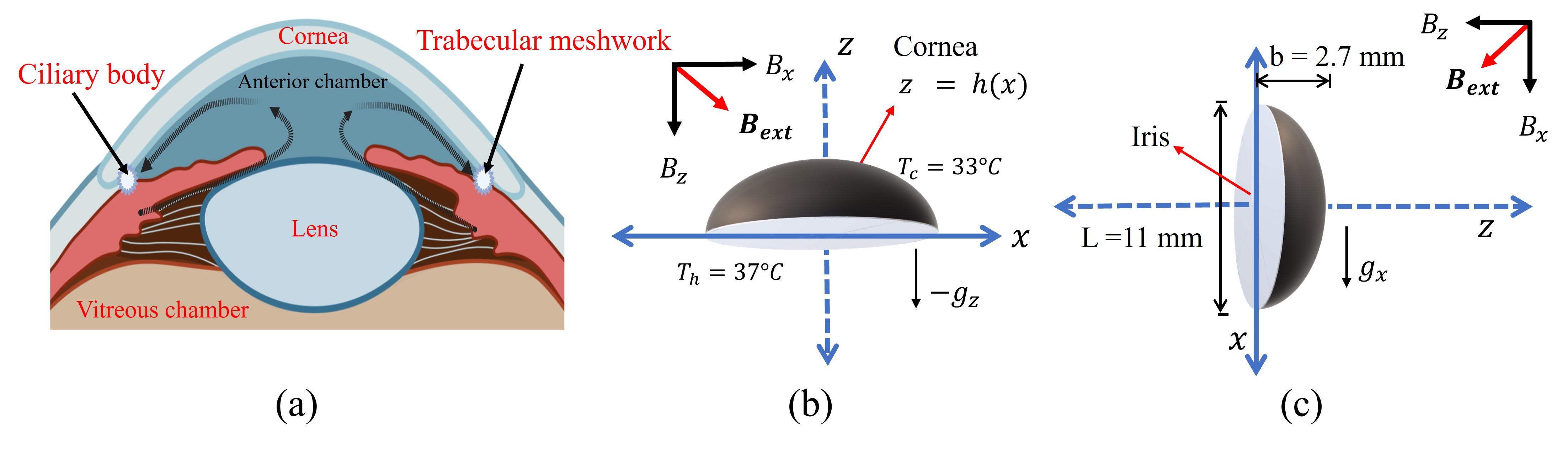}
		\caption{\label{FIG. 1}(a) Schematic of the pathway of aqueous humor from secretion at ciliary body to drainage through trabecular meshwork (created with BioRender.com). Idealization of geometry of anterior chamber of eye in (b) supine position (c)  standing position in the presence of external magnetic field \bm{ $B_{ext} $}.}
	\end{figure*}
	
	Aqueous humor (AH) is continuously produced by the ciliary body in our eye. This fluid enters the anterior chamber through the pupil. It nourishes ocular parts and drains out through Trabecular meshwork as shown in FIG. \ref{FIG. 1}(a). The gap between iris and lens is very small in a healthy eye. The inflow and outflow rates are balanced to maintain constant intraocular pressure and hence the amount of AH within the AC remains constant. The domain considered for the model has iris as a boundary located at $z=0$ and the cornea at $z=h(x)$ (see FIG. \ref{FIG. 1}(b)-(c)).  These are assumed to be solid boundaries (see Appendix \ref{Appendix C}). Our focus is on understanding the flow which is typically symmetric. Hence, we restrict ourselves to a 2D system to generate physical insights. The flow of blood in blood vessels of iris maintains temperature of 37 \degree C at $z=0$. On the other hand, cornea is an avascular tissue and hence it does not contain blood vessels. The corneal temperature (at $z=h(x)$) remains lower (around 33 \degree C) aided by the blinking of the eye and continuous evaporation of the tear film.  This temperature gradient and the small dimension along the $z$-direction induce natural convection and circulation of the AH. FIG. \ref{FIG. 1}((b)-(c)) represents the anterior chamber of eye in the presence of an external magnetic field \bm{$B_{ext} $}. In this work we neglect the inflow and outflow of the AH in the AC.
	We use the Boussinesq approximation and consider that the fluid density varies linearly with temperature only in the gravitational term (equation \ref{eq:1}). Thus, the density of AH is given as
	\begin{equation} \label{eq:1}
		\rho = \rho_{0}\left(1-\alpha(T-T_c)\right).
	\end{equation}
	where $\rho_0$ represents the density of AH at reference temperature $T_c$ and $\alpha$ is the thermal expansion coefficient of AH. AH is considered as an incompressible Newtonian fluid. The flow of AH in the AC is governed by mass, momentum and energy balance equations represented vectorially as
	\begin{equation} \label{eq:2}
		\nabla \cdot \boldsymbol{v}=0,
	\end{equation}
	\begin{equation} \label{eq:3}
		\rho_0\left(\frac{\partial \boldsymbol{v}}{\partial t}+\boldsymbol{v} \cdot \nabla \boldsymbol{v}\right)=-\nabla p+\mu \nabla^2 \boldsymbol{v} +\rho \boldsymbol{g}+\boldsymbol{J} \times \boldsymbol{B}_{\text {ext }},
	\end{equation}
	\begin{equation} \label{eq:4}
		\rho_0 c_p\left(\frac{\partial T}{\partial t}+\boldsymbol{v} \cdot \nabla T\right)=k \nabla^2 T+\mu \phi_h+\frac{\boldsymbol{J}^2}{\sigma},
	\end{equation}
	
	where $\boldsymbol{J}=\sigma(\boldsymbol{v} \times \boldsymbol{B_{ext}}), \nabla \equiv (\frac{\partial}{\partial x},\frac{\partial}{\partial z}), \nabla^2 \equiv (\frac{\partial^2}{\partial x^2},\frac{\partial^2}{\partial z^2}). $\\

	The velocity vector is denoted as $\boldsymbol{v}$ with velocity components $(u,0,w)$ in $x$, $y$ and $z$ directions. $\boldsymbol{J}$ is the current density and $\boldsymbol{B}_{\text{ext}} = (B_x, 0, B_z)$ is external magnetic flux density with $B_x$ and $B_z$  as its components along the $x$ and $z$ axes respectively. The term $\boldsymbol{J} \times \boldsymbol{B}_{\text{ext}}$ represents the magnetic body force per unit volume in equation (\ref{eq:3}). The symbols $\mu$ and $c_p$ denote the viscosity and specific heat of AH. $\phi_h$ and $\frac{\boldsymbol{J}^2}{\sigma}$ represent the dissipation of energy due to viscosity and magnetic field respectively. The direction of acceleration due to gravity $\boldsymbol{g}$ depends on the position of eye. In standing position, gravity acts in positive $x$-direction as shown in FIG \ref{FIG. 1}(c). Conversely, gravity acts in negative $z$-direction in supine position as shown in  FIG. \ref{FIG. 1}(b). We choose following dimensionless variables to nondimensionalize the governing equations (\ref{eq:2})-(\ref{eq:4})\\
	$x^* = \frac{x}{L}, \hspace{1pt} z^* = \frac{z}{\epsilon L},\hspace{1pt} u^* = \frac{u}{U},\hspace{1pt} w^* = \frac{w}{\epsilon U},\hspace{1pt} t^* = \frac{t U}{L},\hspace{1pt} p^* = \frac{p \epsilon^2 L}{\mu U},\hspace{1pt}	B_i^* = \frac{B_i}{B_0},\hspace{1pt} T^* = \frac{T - T_c}{T_h - T_c},$
	where $\epsilon = \frac{b}{L},\hspace{1pt} U = \frac{\mu}{\rho_0 L}$ and $i = (x,y)$\\
	
	This results in the dimensionless governing equations 
	\begin{equation}\label{eq:5}
		\frac{\partial u^*}{\partial x^*} + \frac{\partial w^*}{\partial z^*} = 0,
	\end{equation}
	
	\begin{equation}\label{eq:6}
		\begin{split}
			\epsilon^2 Re \left( \frac{\partial u^*}{\partial t^*} + u^* \frac{\partial u^*}{\partial x^*} + w^* \frac{\partial u^*}{\partial z^*} \right) = -\frac{\partial p^*}{\partial x^*} + \epsilon^2 \frac{\partial^2 u^*}{\partial {x^*}^2} + \frac{\partial^2 u^*}{\partial {z^*}^2}\\ + \frac{\epsilon^2 Re}{Fr_x} \left( 1 - \alpha (T_h - T_c) T^* \right)\\ + \epsilon^2 Ha^2 \left( \epsilon B_x^* B_z^* w^* - {B_z^*}^2 u^* \right),
		\end{split}
	\end{equation}
	
	\begin{equation}\label{eq:7}
		\begin{split}
			\epsilon^4 Re\left(\frac{\partial w^*}{\partial t^*} + u^* \frac{\partial w^*}{\partial x^*} + w^* \frac{\partial w^*}{\partial z^*}\right) = -\frac{\partial p^*}{\partial z^*} + \epsilon^4 \frac{\partial^2 w^*}{\partial {x^*}^{2}}\\ + \epsilon^2 \frac{\partial^2 w^*}{\partial {z^*}^{2}} + \frac{\epsilon^3 Re}{Fr_z} (1-\alpha (T_h-T_c) T^*) +\\ \epsilon^3 Ha^2 (B_x^* B_z^* u^* - \epsilon {B_x^*}^{2} w^*),
		\end{split}
	\end{equation}
	
	\begin{equation}\label{eq:8}
		\begin{split}
			\frac{\partial T^*}{\partial t^*} + u^* \frac{\partial T^*}{\partial x^*} + w^* \frac{\partial T^*}{\partial z^*} =\frac{1}{Re Pr} \left( \frac{\partial^2 T^*}{\partial {x^*}^{2}} + \frac{1}{\epsilon^2} \frac{\partial^2 T^*}{\partial {z^*}^{2}} \right)\\ + \frac{2 Br}{\epsilon^2 Re Pr} \Biggl( \epsilon^2 \Bigg(\left( \frac{\partial u^*}{\partial x^*} \right)^2 + \left( \frac{\partial w^*}{\partial z^*} \right)^2 \Bigg) \\+ \frac{1}{2} \left( \frac{\partial u^*}{\partial z^*} + \epsilon^2 \frac{\partial w^*}{\partial x^*} \right)^2\Biggl) + \frac{\epsilon^2 Ha^2 Br}{Re Pr} \left( B_x^* w^* - \frac{B_z^* u^*}{\epsilon} \right)^2.
		\end{split}
	\end{equation}
	The dimensionless parameters appearing in equations (\ref{eq:5})-(\ref{eq:8}) are Reynolds number $Re = \frac{{\rho_0 UL}}{{\mu}}$, Froude number $Fr_i = \frac{{U^2}}{{Lg_i}}$, Prandtl number $Pr = \frac{{\mu c_p}}{{k}}$, Brinkmann number $Br = \frac{{\mu U^2}}{{k(T_h - T_c)}}$ , and Hartmann number $Ha = B_0 L \sqrt{{\frac{{\sigma}}{{\mu}}}}$.
	The physical properties of AH are assumed to be the same as that of water \citep{canning2002fluid},\\
	$\rho_0 = 1000 \hspace{2pt}\text{kg/m}^3, \mu = 0.9 \times 10^{-3}\hspace{2pt} \text{Pa} \cdot \text{s},\alpha = 3 \times 10^{-4}\hspace{2pt} \text{K}^{-1} $\\
	$k = 0.57 \hspace{2pt} \text{W/m} \cdot \text{K}, c_p = 4.2 \times 10^3 \hspace{2pt}\text{J/kg} \cdot \text{K}$\\
	
	The electrical conductivity of AH depends on the presence of various ions and amino acids \citep{goel2010aqueous}. This has been reported as $O(1)$ for AH \citep{lee2022vivo,lee2024multi}. This is dependent on several physiological factors such as ionic concentration, protein content, temperature, and age. Hence, we use $\sigma=5$ S/m in our work.  A magnetic field of $15.2$ T was used for scanning olfactory bulbs in rats \citep{chitrit2023functional}. At this intensity, all the rats recovered after the experiment. A magnetic field gradient of $20$ T/m was found sufficient to help magnetic nanoparticles penetrate the vitreous body \citep{zahn2020investigation}.  Following this, we have taken a magnetic field of $20$ T for our calculation. The depth of anterior chamber $(b)$ is much smaller than its width $(L)$. For human eye, their typical values are taken as $b=2.7$ mm and $L=11$ mm as shown in FIG. \ref{FIG. 1}(c). The values of the dimensionless variables appearing in equations (6)-(8) are given in TABLE \ref{table:dimensionless_numbers}. Using these values, we determine\\\\
	\begin{table}[h!]
		\centering
		\begin{tabular}{cc}
			\textbf{Dimensionless numbers} & \qquad \textbf{Values} \\
			Reynolds number $(Re)$ & \qquad 1 \\
			Prandtl number $(Pr)$ & \qquad 6.63 \\
			Brinkmann number $(Br)$ &  \qquad $2.64 \times 10^{-12}$ \\
			Hartmann number $(Ha)$ & \qquad 16.4 \\
			Froude number $(Fr)$ & \qquad $6.21 \times 10^{-8}$ \\
		\end{tabular}
		\caption{The dimensionless numbers based on the geometrical and physical parameters.}
		\label{table:dimensionless_numbers}
	\end{table}
	
	\begin{equation}
	\epsilon^2 Re \sim 0.06,  \frac{{2Br}}{{\epsilon^2 Re \cdot Pr}} \sim 1.3 \times 10^{-11}, \frac{{\epsilon^2 Ha^2 Br}}{{Re Pr}} \sim 6.4 \times 10^{-12}
	\end{equation}
	\begin{equation} \label{eq:dim}
 	\frac{{\epsilon^2 Re}}{{Fr_i}} \alpha(T_h - T_c) \sim 1.2 \times 10^3, \epsilon^2 Ha^2 \sim 16.2
	\end{equation}

	Based on these values, we conclude that the dissipation due to viscosity and magnetic field can be neglected. At the leading order, the inertial forces are also neglected. From equation (\ref{eq:dim}), we observe that the buoyancy is much higher than the magnetic force. The ratio of magnetic force to the buoyancy is 
	\begin{equation} \label{eq:eta}
		\eta = \frac{F_M}{F_B} = \frac{Ha^2 Fr_i}{Re\alpha(T_h-T_c)}\sim0.014
	\end{equation}
	Low value of $\eta$  indicates that buoyant forces dominate the magnetic forces. Substituting the expressions for the dimensionless numbers in equation (\ref{eq:eta}), we obtain
	\begin{equation}
	\eta = \frac{\sigma B_0^2 /\mu}{\rho_0^2 g \alpha L (T_h-T_c)/\mu^2} = \frac{\lambda_1}{\lambda_0}
	\end{equation}
	Here, $\lambda_1 L^2$ is the ratio of magnetic force and viscous forces whereas $\lambda_0 L^2$ represents ratio of buoyancy to the viscous force. The Rayleigh number based on the depth of the anterior chamber $(b)$ is 1895. This exceeds the critical Rayleigh number $(Ra_c=1700)$  required for the onset of natural convection between two flat solid plates. Hence, we expect natural convection within the anterior chamber of the eye.
	The dimensional governing equations are given as
	
	\begin{equation}\label{eq:9}
		\frac{{\partial u}}{{\partial x}} + \frac{{\partial w}}{{\partial z}} = 0,
	\end{equation}
	\begin{equation}\label{eq:10}
		-\frac{{\partial p}}{{\partial x}} + \mu \frac{{\partial^2 u}}{{\partial z^2}} + \rho_0 g_x \left(1 - \alpha(T - T_c)\right) + \sigma\left(wB_x B_z - B_z^2 u\right) = 0,
	\end{equation}
	\begin{equation}\label{eq:11}
		-\frac{{\partial p}}{{\partial z}} + \rho_0 g_z \left(1 - \alpha(T - T_c)\right) + \sigma\left(uB_x B_z - B_x^2 w\right) = 0,
	\end{equation}
	\begin{equation}\label{eq:12}
		\frac{{\partial^2 T}}{{\partial z^2}} = 0.
	\end{equation}
	
	The above equations are subject to the no-slip and no penetration boundary conditions at the iris and cornea that act as solid boundaries where the temperatures are specified.
	\begin{equation}\label{eq:13}
		u =w = 0 \hspace{5pt} \text{and} \hspace{5pt} T = T_h \hspace{5pt} \text{at} \hspace{5pt} z=0,
	\end{equation}
	\begin{equation}\label{eq:14}
		u = w = 0 \hspace{5pt} \text{and} \hspace{5pt} T = T_c \hspace{5pt} \text{at} \hspace{5pt}  z=h(x).
	\end{equation}
	Here, $z = h(x)$ is the corneal surface. The depth of anterior chamber is maximum at the centre, therefore it is expressed as $z = h(x) = b\sqrt{1 - \frac{x^2}{a^2}}.$
	
	\section{\label{secIII}Results and discussion}
	\subsection{Influence of uniform magnetic field in $z$-direction on AH dynamics }
	\subsubsection*{A.1 Supine position}
	$\left.\begin{array}{l} \boldsymbol{g} = (0,-g), \boldsymbol{B}_{\text{ext}} = (0,B_z) \end{array}\right.$\\
	
	In supine position, a person faces vertically upward and gravity acts in negative $z$-direction as shown in FIG. \ref{FIG. 1}(b). Here, we consider a constant magnetic flux density $(B_z)$ in negative $z$-direction. The behaviour of the system is independent of whether the magnetic field is in positive or negative $z$- direction since its contribution arises as $B_z^2$. The temperature of the AH is determined by a linear equation (\ref{eq:12}) subject to equations (\ref{eq:13})-(\ref{eq:14}). The solution is given by 
	
	\begin{figure}
		\includegraphics[width=0.49\textwidth]{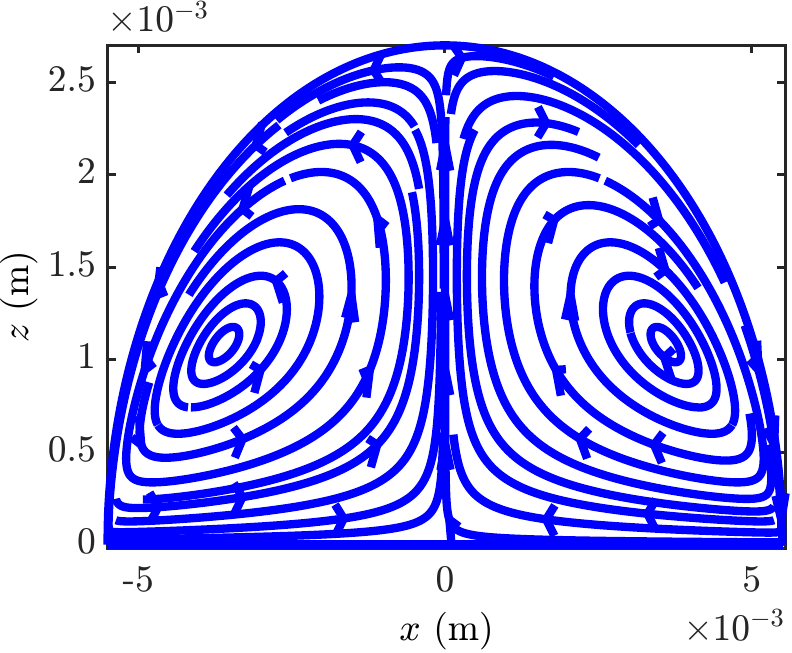}
		\caption{\label{FIG. 2}Streamlines of AH flow under the influence of uniform magnetic field $B_z=20$ T in supine position.}
	\end{figure}

	\begin{equation} \label{eq:15}
		T = T_h - \left( T_h - T_c \right) \frac{z}{h(x)}.
	\end{equation}
	We eliminate pressure from equations (\ref{eq:10}) and (\ref{eq:11}) to obtain 
	\begin{equation}\label{eq:16}
		\frac{\partial^3 u}{\partial z^3}  - \lambda_1 \frac{\partial u}{\partial z} - \lambda \frac{z\hspace{1pt} h'(x)}{h(x)^2} = 0,
	\end{equation}
	
	where $\lambda = \frac{\rho_0 g \alpha (T_h-T_c)}{\mu}$ and $\lambda_1 = \frac{\sigma B_z^2}{\mu}$. The equation (\ref{eq:16}) is rescaled using $u=\tilde u/\lambda_{0}$. The results
	\begin{equation}\label{eq:17}
		\frac{1}{\lambda_0}  \frac{\partial^3 \tilde u}{\partial z^3}- \eta \frac{\partial \tilde u}{\partial z} - \lambda \frac{z h'(x)}{h(x)^2} = 0.
	\end{equation}
 	This rescaling explicitly shows the magnetic forces (second term) are relatively small  compared to buoyancy forces.  In the limit $\eta\rightarrow0$, this reduces to purely buoyancy-driven flow. This rescaling helps us obtain the effect of magnetic field using a perturbation analysis. We seek solution in the form of a perturbation series in small parameter $\eta$ as
	\begin{equation}\label{eq:18}
		\tilde u = \tilde u_0 + \eta \tilde u_1 + \eta^2 \tilde u_2 + \cdots
	\end{equation}
	\begin{equation}\label{eq:19}
		\tilde	w =\tilde w_0 + \eta \tilde w_1 + \eta^2 \tilde w_2 + \cdots
	\end{equation}
	\begin{figure*}
		\includegraphics[width=1\textwidth]{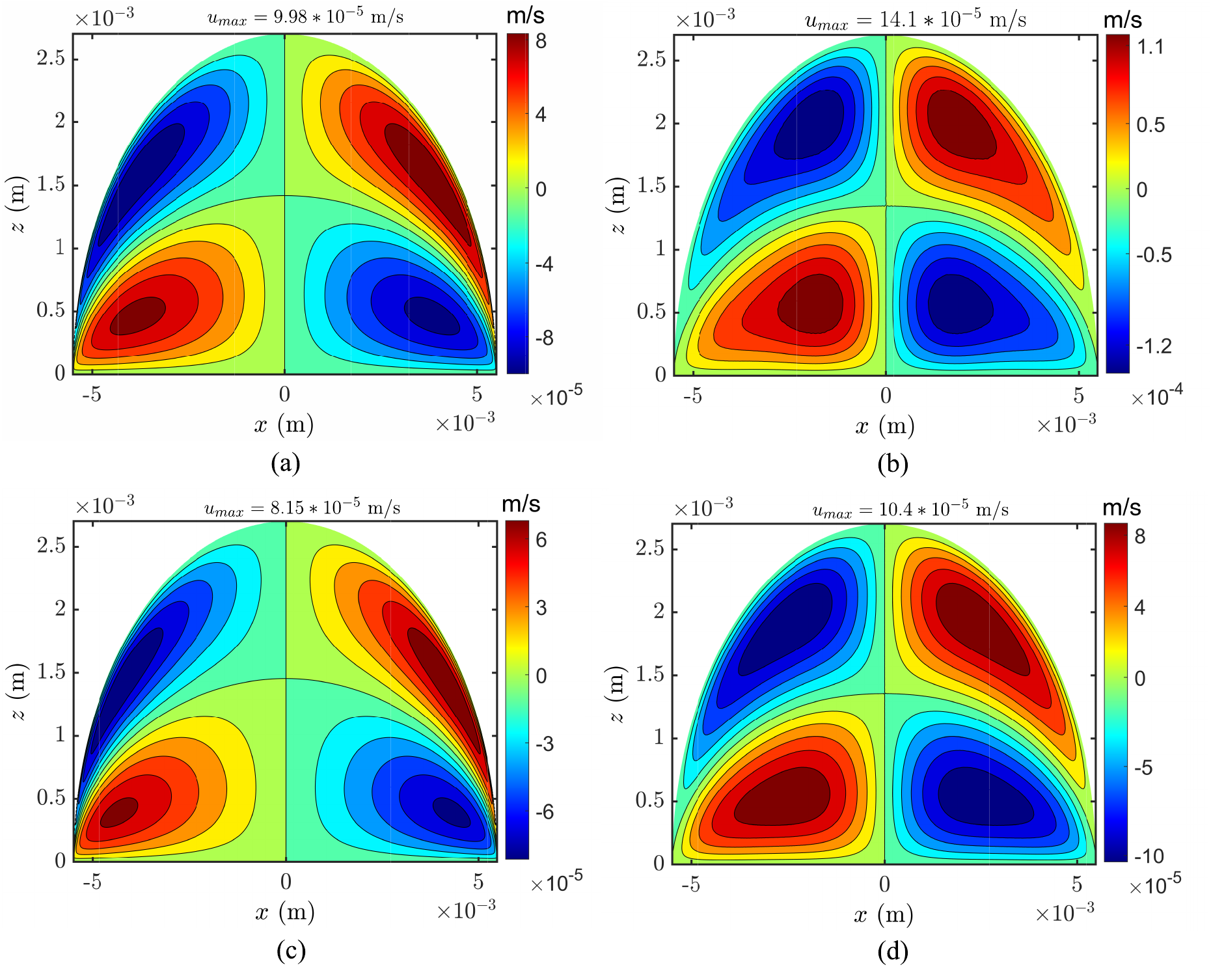}
		\caption{\label{FIG. 3}Contour plots of $x$-component of AH velocity in supine position  (a) analytically obtained velocity $(u_0)$ solely due to buoyancy $(\eta = 0)$ (b) numerically obtained velocity $(u_0)$ solely due to buoyancy $(\eta = 0)$ (c) analytically obtained velocity $(u)$ considering both buoyancy and magnetic field $(B_z=20$ T, $\eta = 0.014)$ (d) numerically obtained velocity $(u)$ considering both buoyancy and magnetic field $(B_z=20$ T, $\eta = 0.014)$.}
	\end{figure*}
	The unperturbed velocity components denoted by $\tilde u_0$ and $\tilde w_0$  represent the velocity of AH solely due to buoyancy. $\tilde u_1$ and $\tilde w_1$ represent the corrections in velocity due to presence of external magnetic field. Collecting terms at different order of $\eta$, we obtain
	\begin{equation}\label{eq:20}
		O(\eta^0) : \frac{1}{\lambda_0}  \frac{\partial^3 \tilde u_0}{\partial z^3}  - \lambda \frac{zh'(x)}{h(x)^2} = 0.
	\end{equation}
	From equation of continuity, we obtain
	\begin{equation}\label{eq:21}
		\frac{\partial \tilde u_0}{\partial x} + \frac{\partial \tilde w_0}{\partial z} = 0.
	\end{equation}
	
	\begin{equation}\label{eq:22}
		O(\eta^1) : \frac{1}{\lambda_0}  \frac{\partial^3 \tilde u_1}{\partial z^3}  -  \frac{\partial \tilde u_0}{\partial z} = 0.
	\end{equation}
	From equation of continuity, we obtain
	\begin{equation} \label{eq:23}
		\frac{\partial \tilde u_1}{\partial x} + \frac{\partial \tilde w_1}{\partial z} = 0.
	\end{equation}
	
	The corresponding boundary conditions are 
	\begin{equation} \label{eq:24}
		\tilde{u}_0 = \tilde{w}_0 = \tilde{u}_1 = \tilde{w}_1 = 0 \quad \text{at} \quad z=0 \quad \text{and} \quad z=h(x).
	\end{equation}
	Integrating equation (\ref{eq:20}) with respect to $z$, we obtain 
	\begin{equation}\label{eq:25}
		\tilde{u}_0 = \frac{\lambda \lambda_0 h'(x) z^4}{24 h(x)^2} + c_1(x) \frac{z^2}{2} + c_2(x) z + c_3(x),
	\end{equation}
	where $c_1 (x),c_2 (x)$ and $c_3 (x)$ are to be determined using boundary conditions. The no slip at iris $(z=0)$, yields
	\begin{equation}\label{eq:26}
		c_3 (x)=0.
	\end{equation}
	The no slip boundary condition at the corneal surface $(z=h(x))$ gives 
	\begin{equation}\label{eq:27}
		c_2(x) = -\frac{\lambda\lambda_0 h'(x) h(x)}{24} - \frac{c_1(x) h(x)}{2}
	\end{equation}
			\begin{figure*}
		\includegraphics[width=1\textwidth]{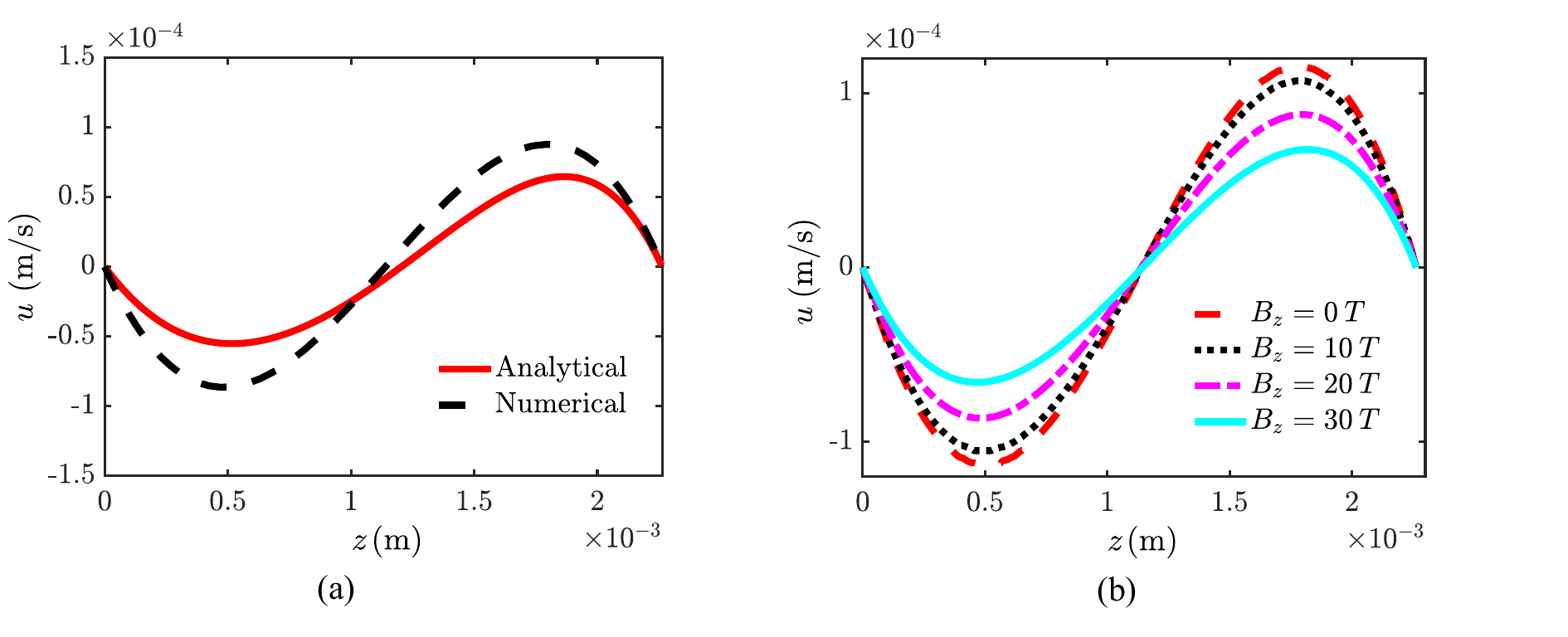}
		\caption{\label{FIG. 4}(a) Comparison between the analytically and numerically obtained velocity $(u)$ for supine position in the presence of uniform magnetic flux density $B_z=20$ T at $x=3$ mm  (b) Effect of magnetic field on the velocity $(u)$ within the anterior chamber obtained at $x=3$ mm.}
	\end{figure*}
	
	\begin{figure*}
		\includegraphics[width=1\textwidth]{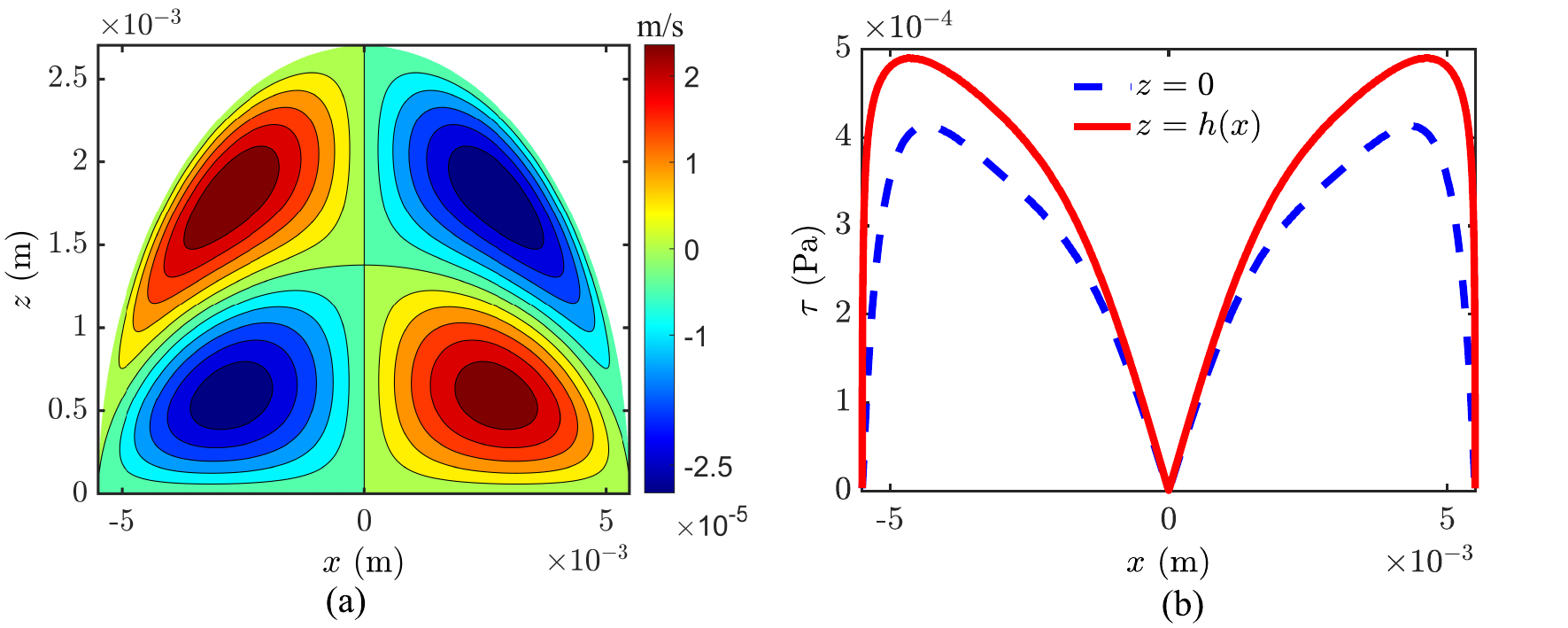}
		\caption{ \label{FIG. 5}(a) Contour plot of $x$-component of velocity $u_1$ solely due to presence of magnetic field $B_z=20$ T (b) Shear stress on iris $(z=0)$  and cornea $(z=h(x))$ in the presence of uniform magnetic field $B_z=20$ T in supine position.}
	\end{figure*}
	We obtain $\tilde{w}_0$ using continuity equation (\ref{eq:21}) as
	\begin{equation}\label{eq:28}
		\begin{split}
			\tilde	w_0 = -\left(\frac{h''(x)h(x)-2h'(x)^2}{h(x)^3}  \frac{\lambda \lambda_0 z^5}{120} + \frac{c_1'(x)z^3}{6}\right)-\\ \frac{c_2'(x)z^2}{2} - c_4(x).
		\end{split}
	\end{equation}
	The no penetration condition at iris, $z=0$ gives
	\begin{equation}\label{eq:29}
		c_4 (x)=0.
	\end{equation}
	The no penetration condition at cornea, $z=h(x)$ and equation (\ref{eq:27}) yield 
	\begin{equation}\label{eq:30}
		2c_1'(x)h(x) + 6c_1(x)h'(x) = -\frac{3\lambda \lambda_0}{10} \left(3h'(x)^2 + h(x)h''(x)\right).
	\end{equation}
	$c_1 (x)$ is obtained by solving equation (\ref{eq:30}). This yields
	\begin{equation}\label{eq:31}
		c_1(x) = \frac{3b^2 \lambda\lambda_0 x}{20a^2 h(x)} + \frac{c_5}{(a^2-x^2)^{3/2}},
	\end{equation}
	where $c_5$ is an arbitrary constant. To find $c_5$, we use the fact that pressure is a point function and thus the change in pressure over any closed surface should be zero \citep{ramji2019modelling}. Hence,
	\begin{equation}\label{eq:32}
		\oint  \frac{\partial p}{\partial s}\, ds =0.
	\end{equation}
		
\begin{figure*}
	\includegraphics[width=1\textwidth]{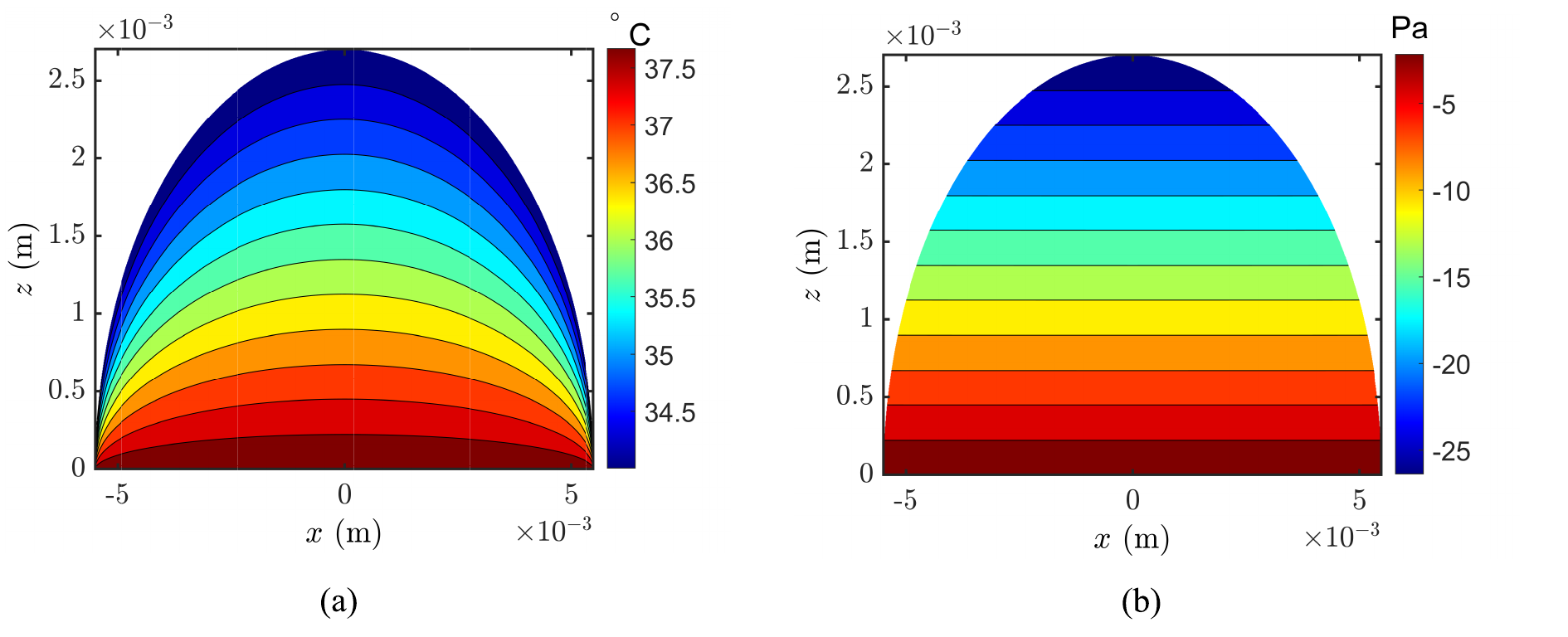}
	\caption{ \label{FIG. 6}Contour plot of (a) temperature of AH (b) pressure of AH in anterior chamber of eye in supine position.}
\end{figure*}	
		
	The evaluation of the integral and pathway of the closed curve has been given in FIG. \ref{FIG. A.1} of Appendix \ref{Appendix A}. This gives $c_5=0$. Hence
	\begin{flalign}\label{eq:33}
		\tilde u_0 = &\lambda\lambda_0 \bigg[\frac{h'(x)z^4}{24h(x)^2} + \frac{3b^2 x}{20a^2 h(x)}\frac{z^2}{2} - \bigg(\frac{3b^2 x}{40 a^2}\nonumber\\
		& + \frac{h(x)h'(x)}{24}\bigg)z\bigg],
	\end{flalign}
	
	\begin{flalign}\label{eq:34}
		\tilde w_0 = &-\lambda \lambda_0 \bigg[ \frac{h''(x)h(x)-2h'(x)^2}{h(x)^3}\frac{z^5}{120} + \frac{3b^2(h(x)-xh'(x))}{20a^2 h(x)^2} \frac{z^3}{6} \nonumber\\ &- \frac{9b^2+5a^2(h'(x)^2+h(x)h''(x))}{120a^2}\frac{z^2}{2}\bigg].
	\end{flalign}
	
	From equation (\ref{eq:22}) we obtain 
	\begin{equation}\label{eq:35}
		\begin{split}
			\tilde u_1 = \frac{1}{720} z^3 \lambda_0\left( 30z c_1(x) + 120 c_2(x) + \frac{z^3 \lambda\lambda_0 h'(x)}{h(x)^2}\right)\\ + A_1(x) z^2 + A_2(x) z + A_3(x).
		\end{split}
	\end{equation}
	The arbitrary constants $A_1 (x),A_2 (x),A_3 (x)$ are obtained using the (equations (\ref{eq:13})-(\ref{eq:14})). From no slip condition, we obtain 
	\begin{equation}\label{eq:36}
		A_3 (x)=0,
	\end{equation}
	\begin{flalign}\label{eq:37}
		A_2(x) =& -\frac{1}{720} h(x)\bigg(720 A_1(x) + \nonumber \\ & \lambda_0 h(x)\bigg(120 c_2(x) + h(x)(30 c_1(x) + \lambda \lambda_0 h'(x)\bigg)\bigg).
	\end{flalign}
	Using continuity equation we obtain,
	\begin{equation}\label{eq:38}
		\begin{split}
			\tilde w_1 = A_4(x) - \frac{1}{720 h(x)^3} \bigg(240z^3 h(x)^3 A_1'(x) +\\ 360z^2 h(x)^3 A_2'(x) + 720zh(x)^3 A_3'(x) +\\ 6z^5 \lambda_0 h(x)^3 c_1'(x) + 30z^4 \lambda_0 h(x)^3 c_2'(x)\\ - \frac{2}{7} z^7 \lambda\lambda_0^2 h'(x)^2 + \frac{1}{7} z^7 \lambda\lambda_0^2 h(x) h''(x)\bigg).
		\end{split}
	\end{equation}
	Using no penetration condition at $z=0$ and $z=h(x)$ yields
	\begin{equation}\label{eq:39}
		A_4 (x)=0.
	\end{equation}
	The differential equation for $A_1' (x)$ is given by 
	\begin{equation}\label{eq:40}
		\begin{split}
			5040A_1(x) h'(x) + 1680h(x)(A_1'(x) + \lambda_0 c_2(x) h'(x)) \\+ 5\lambda_0 h(x)^2 (84c_2'(x) + h'(x)(126c_1(x) + 5\lambda\lambda_0 h'(x))\\ + \lambda_0 h(x)^3 (126c_1'(x) + 5\lambda\lambda_0 h''(x)) = 0.
		\end{split}
	\end{equation}
	Solving equation (\ref{eq:40}), we obtain
	\begin{equation}\label{eq:41}
		\begin{split}
			A_1(x) = \frac{A_5}{h(x)^3} -  \frac{1}{4} \lambda_0 c_2(x)h(x)\\ - \frac{\lambda_0 h(x)^2 (126c_1(x) + 5\lambda\lambda_0 h'(x))}{1680}.
		\end{split}
	\end{equation}
	
	From equation (\ref{eq:32}), we obtain that $A_5=0$. The velocities are rescaled back into its original form using equations (\ref{eq:18})-(\ref{eq:19}) to obtain
	\begin{equation}\label{eq:42}
		u = \frac{1}{\lambda_0} (\tilde u_0 + \eta \tilde u_1) = u_0 + u_1 \quad \text{and} \quad	w = \frac{1}{\lambda_0} (\tilde w_0 + \eta \tilde w_1) = w_0 + w_1
	\end{equation}
	In the above, $u_0=\frac{\tilde u_0}{\lambda_0}$  represents the velocity due to natural convection effects alone and $u_1= \frac{\eta \tilde u_1}{\lambda_0}$  represents the velocity correction due to the magnetic field.

	FIG. \ref{FIG. 2} represents the streamlines of the AH flow in the anterior chamber. The warmer fluid near the iris rises at the center and moves towards cornea. The flow is symmetric about $x=0$, resulting in two counter rotating vortices.\\
	We have also performed numerical simulation using COMSOL Multiphysics 6.2\textsuperscript \textregistered  to validate the accuracy of our analytical solution. The full set of governing equations (\ref{eq:2})-(\ref{eq:4}) are solved using stationary solver along with PARADISO and a fully coupled solver in COMSOL. A maximum relative tolerance of $10^{-7}$ is applied to all dependent variables for convergence. We employ a user-controlled meshes with an extremely fine triangular finite elements selecting a minimum element size of $1.05*10^{-7}$ m. A grid independency analysis was performed by calculating the velocity of the AH at $x = 3$ mm for different numbers of grid elements. The details of these grid elements have given in the Supplementary material (section S.1). The velocity does not change once the number of domain elements exceeds 24,662 as shown in FIG. S1 in the Supplementary material. Hence, we choose this grid for further computations. The reference pressure point is selected as zero at the center of iris.\\
	FIG. \ref{FIG. 3}(a) depicts the contour plot of $u_0=\frac{\tilde u_0} {\lambda_0}$ obtained analytically (equation (\ref{eq:33})) using lubrication approximation.FIG. \ref{FIG. 3}(a) depicts the contour plot of $u_0=\frac{\tilde u_0} {\lambda_0}$ obtained analytically (equation (\ref{eq:33})) using lubrication approximation. FIG. \ref{FIG. 3}(b) represents the corresponding plot obtained from numerical simulation. We see that the two solutions are in good agreement. FIG. \ref{FIG. 3}(c) shows $x$-component of velocity $(u)$ obtained analytically considering buoyancy and magnetic field effects. Lorentz force is a body force and it opposes buoyancy in this case.  As a result, the net driving force responsible for fluid motion decreases, leading to a reduction in velocity. This is seen on comparing FIG. \ref{FIG. 3}(c) with FIG. \ref{FIG. 3}(a). FIG. \ref{FIG. 3}(d) shows the contour plot of  $(u)$ obtained numerically. The velocity profiles obtained analytically and numerically are in good agreement. FIG. \ref{FIG. 3}(b) reveals that the maximum magnitude of AH velocity, driven solely by buoyancy, reaches $14.1*10^{-5} \text{m/s}$. This reduces to $10.4*10^{-5}  \text{m/s}$ in the presence of a magnetic field. Therefore, a uniform magnetic field of $20 \hspace{1pt} \text{T}$ directed along $z$-direction leads to a 26.24 \% decrease in AH velocity. 
	
	\begin{figure}
		\includegraphics[width=0.45\textwidth]{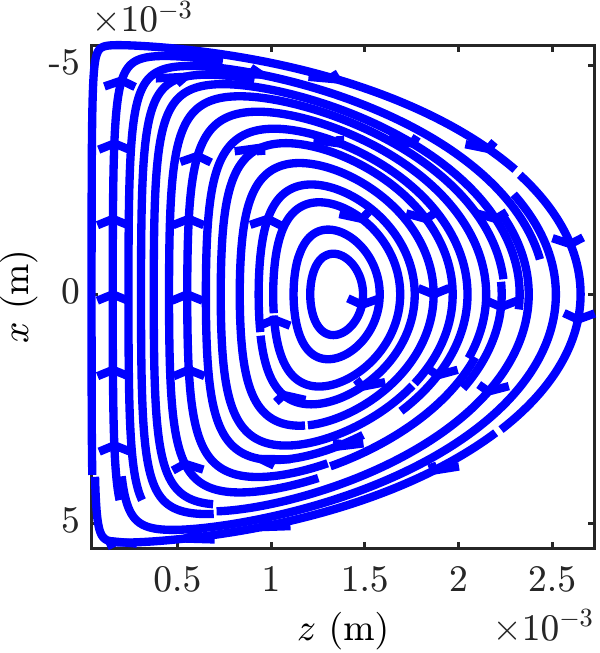}
		\caption{\label{FIG. 7}Streamlines of AH flow in the presence of uniform magnetic field $B_z=20$ T in standing position.}
	\end{figure}
	To verify the quantitative match between the analytical and numerical solution, we plot variation $u$ with respect to $z$ at $x=3$ mm (which is almost at the middle of right half of the anterior chamber) in FIG. \ref{FIG. 4}(a). We see that though $\epsilon$ is not too low, the analytical solution matches well with the numerical solution. The agreement between these two solutions improves significantly for smaller $\epsilon$ values. They match completely for $\epsilon=0.045$ (see Supplementary material FIG. S2 (a)).  FIG. \ref{FIG. 4}(b) illustrates the effect of magnetic field on the velocity $(u)$ of AH at $x=3$ mm. As the strength of the uniform magnetic field increases, the opposing body force $\sigma B_z^2 u$ increases in strength. Consequently, the net driving force decreases leading to a lower AH velocity. Lorentz forces are proportional to square of the magnetic strength, hence the decrease in velocity is more significant for higher magnetic strength values.

	FIG. \ref{FIG. 5}(a) shows the contour plot of $u_1$ induced by magnetic field obtained analytically using equation (\ref{eq:35}). The Lorentz forces generate flow in opposite direction to that generated by buoyancy as it resists the cause of flow \citep{wang2019flow}. This leads to fluid movement in a direction opposite to that driven solely by buoyancy as shown in FIG. \ref{FIG. 5}(a).
	The wall shear stress (WSS) is given by,
	\begin{equation}\label{eq:43}
		\tau = \mathbf{\hat{n}}\cdot\mathbf{S}\cdot\mathbf{\hat{t}}
	\end{equation}
	where $\mathbf{\hat{n}}$ and $\mathbf{\hat{t}}$ are the unit normal and tangent vectors on the wall respectively. $\mathbf{S} = -p\mathbf{I}+2\mu\mathbf{E}$ is total stress tensor with $\mathbf{E}=\frac{1}{2} (\nabla \boldsymbol{v}+\nabla \boldsymbol{v}^T)$. The unit normal and unit tangent vectors to the corneal surface are given by
	\begin{equation}\label{eq:44}
		\mathbf{\hat{n}} = \frac{-h'(x)\mathbf{\hat{e}}_x + \mathbf{\hat{e}}_z}{\sqrt{1 + h'(x)^2}}
	\end{equation}
	\begin{equation}\label{eq:45}
		\mathbf{\hat{t}} = \frac{\mathbf{\hat{e}}_x + h'(x) \mathbf{\hat{e}}_z}{\sqrt{1 + h'(x)^2}}
	\end{equation}
		\begin{figure*}
		\includegraphics[width=1\textwidth]{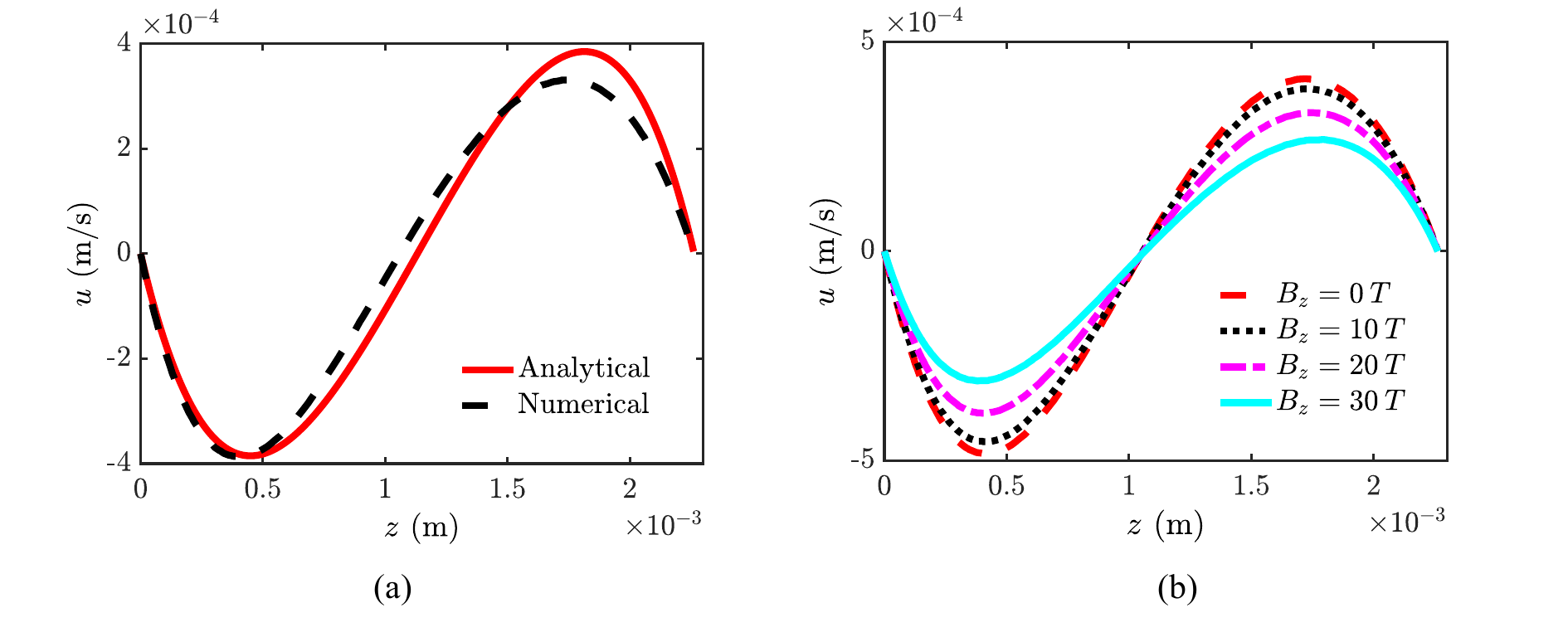}
		\caption{ \label{FIG. 8}(a) Comparison between the analytically and numerically obtained velocity $(u)$ in the presence of uniform magnetic flux density $B_z=20$ T at $x=3$ mm in standing position (b) Effect of magnetic field on the velocity $(u)$ within the anterior chamber at $x=3$ mm in standing position.}
		\end{figure*}
		
		\begin{figure}
			\includegraphics[width=0.45\textwidth]{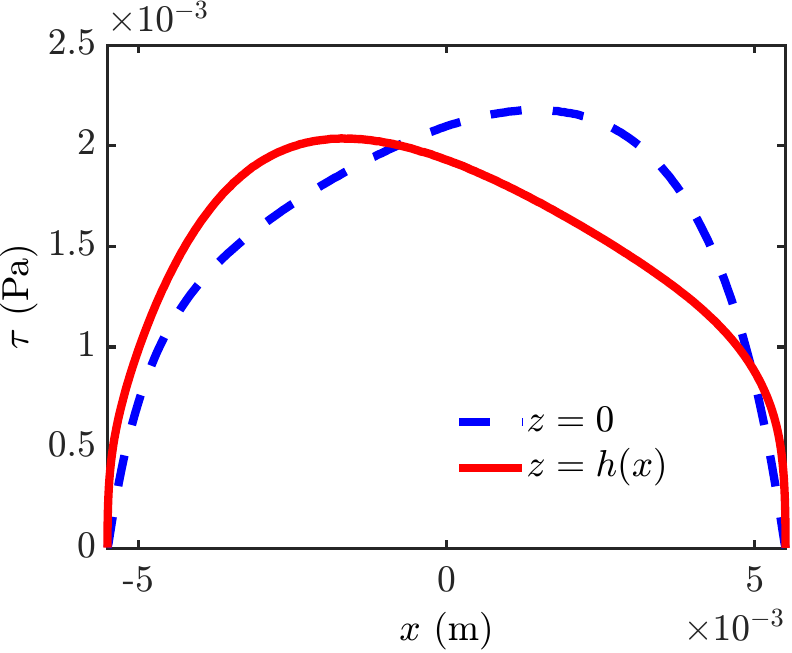}
			\caption{\label{FIG. 9}Shear stress on iris $(z=0)$  and cornea $(z=h(x))$ in the presence of uniform magnetic field $B_z=20$ T in standing position.}
		\end{figure}
	The complete derivation for $\tau$ on corneal and iris surface is given in the Appendix \ref{Appendix C}. FIG. \ref{FIG. 5}(b) represents numerically obtained shear stress on pupil and corneal wall. The cornea experiences a higher shear stress compared to the iris. As the flow is symmetric about $x=0$ (see FIG. \ref{FIG. 2}), the shear stress is also symmetric about $x=0$. The numerically and analytically obtained shear stress profiles are compared in the Supplementary material. As illustrated in FIG. S3(a), they match well for small $\epsilon$ value. The endothelial cells in human eye detach for shear stress greater than $0.03$ Pa \citep{kaji2005effect}. The maximum shear stress on cornea predicted by our model is $4.8*10^{-4}$ Pa. This is sufficiently low and there is no risk of detachment of endothelial cells in supine position.

	FIG. \ref{FIG. 6}(a) depicts the distribution of temperature inside the anterior chamber of eye. The temperature varies linearly in the $z$ direction for all $x$. The temperature of the aqueous humor decreases from bottom to top inducing natural convection of AH. FIG. \ref{FIG. 6}(b) represents the variation of pressure inside the anterior chamber of eye. There is no pressure variation along $x$ direction as predicted by equations (\ref{eq:10})-(\ref{eq:11}). The pressure increases as we move from cornea to iris.
	
	\subsubsection*{A.2 Standing position}
	$\left.\begin{array}{l} \boldsymbol{g} = (g,0), \boldsymbol{B}_{\text{ext}} = (0,B_z) \end{array}\right.$\\
	
	In standing position, a person faces horizontally and gravity acts in positive $x$-direction as shown in FIG. \ref{FIG. 1}(c). We consider constant magnetic flux density $(B_z)$ acting in $z$-direction. Here again the temperature of the AH varies linearly with $z$ as given by equation (\ref{eq:15}). In this case, gravitational force and the Lorentz force appear in the $x$-momentum balance. Eliminating pressure as described earlier we obtain 
		\begin{figure*}
		\includegraphics[width=1\textwidth]{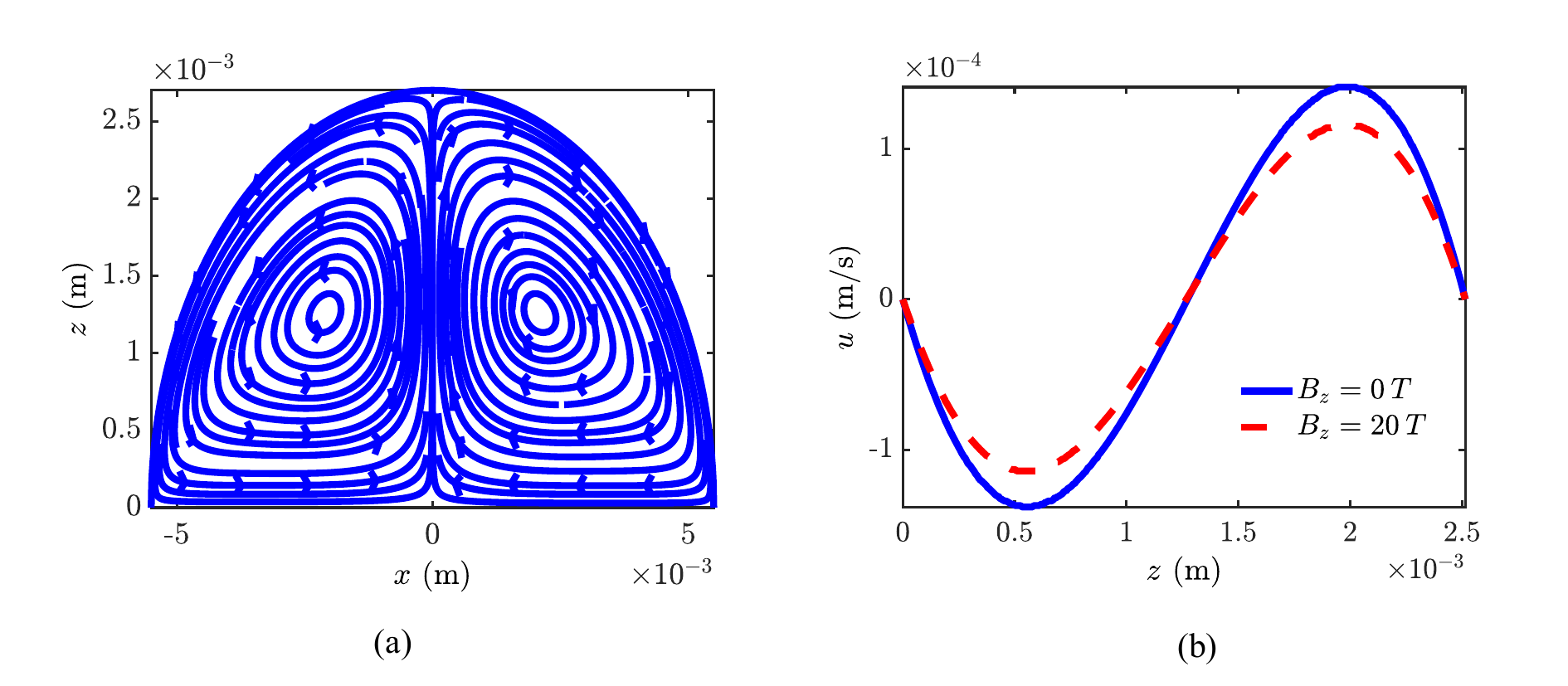}
		\caption{ \label{FIG. 10}(a) Velocity streamlines of AH flow in supine position in the presence of uniform magnetic field $B_x=20$ T (b) Effect of magnetic field on the velocity $(u)$ within the anterior chamber at $x=3$ mm.}
		\end{figure*}
		\begin{figure}
			\includegraphics[width=0.45\textwidth]{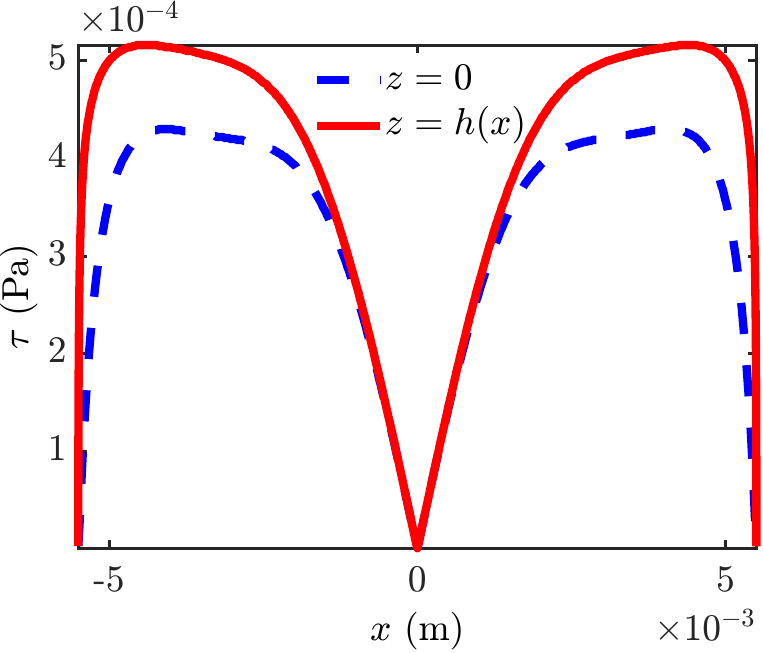}
			\caption{\label{FIG. 11}Shear stress on iris $(z=0)$  and cornea $(z=h(x))$ in the presence of uniform magnetic field $B_x=20$ T in supine position.}
		\end{figure}
	\begin{equation}\label{eq:46}
		\frac{\partial^3 u}{\partial z^3} - \lambda_1 \frac{\partial u}{\partial z} + \frac{\lambda}{h(x)} = 0
	\end{equation}
	where $\lambda = \frac{\rho_0 g \alpha (T_h-T_c)}{\mu}$ and $\lambda_1 = \frac{\sigma B_z^2}{\mu}$.
	We rescale the parameters to obtain a solution using a regular perturbation series expansion as before $u=\tilde u/\lambda_{0}$. Collecting terms at different order of $\eta$, we obtain,\\
	\begin{equation}\label{eq:47}
		O(\eta^0) : \frac{1}{\lambda_0}  \frac{\partial^3 \tilde u_0}{\partial z^3} +  \frac{\lambda}{h(x)} = 0
	\end{equation}
	From equation of continuity, we obtain
	\begin{equation}\label{eq:48}
		\frac{\partial \tilde u_0}{\partial x} + \frac{\partial \tilde w_0}{\partial z} = 0
	\end{equation}

	\begin{equation}\label{eq:49}
		O(\eta^1) : \frac{1}{\lambda_0}  \frac{\partial^3 \tilde u_1}{\partial z^3}  - \frac{\partial \tilde u_0}{\partial z} = 0
	\end{equation}
	From equation of continuity, we obtain
	\begin{equation}\label{eq:50}
		\frac{\partial \tilde u_1}{\partial x} + \frac{\partial \tilde w_1}{\partial z} = 0
	\end{equation}
	This implies 
	\begin{equation}\label{eq:51}
		\tilde u_0 = -\frac{\lambda \lambda_0 z^3}{6h(x)} + c_1(x) \frac{z^2}{2} + c_2(x)z + c_3(x),
	\end{equation}

	\begin{equation}\label{eq:52}
		\tilde w_0 = -\left(\frac{\lambda\lambda_0 z^4 h'(x)}{24h(x)^2} + \frac{c_1'(x)z^3}{6} + \frac{c_2'(x)z^2}{2} + c_4(x)\right)
	\end{equation}
	The expressions $c_1 (x),c_2 (x),c_3 (x),c_4 (x)$ have been given in Appendix \ref{Appendix B}. We substitute the value of $\tilde u_0$ into equation (\ref{eq:22}) and obtain $\tilde u_1$ by integrating this equation with respect to $z$. This gives
	\begin{align}\label{eq:53}
		\tilde{u}_1 = &\ \frac{1}{120}z^3\lambda_0\bigg[5z c_1(x) + 20c_2(x) - \frac{z^2 \lambda \lambda_0}{h(x)}\bigg] \nonumber \\ 
		& + A_1(x)z^2 + A_2(x)z + A_3(x)
	\end{align}
	Using equation of continuity, we obtain
	\begin{align}\label{eq:54}
		\tilde w_1 &= A_4(x) + \frac{1}{720}z^2\bigg[120(-2z+3h(x))A_1'(x) + 3\lambda_0(-2z^3+\\
		& \nonumber 5h(x)^3)c_1'(x) - 30\lambda_0(z^2-2h(x)^2)c_2'(x)+ 15\bigg(24A_1(x) \\
		& \nonumber+\lambda_0h(x)(8c_2(x)+3c_1(x)h(x))\bigg)h'(x) \\
		& \nonumber - \frac{ \lambda \lambda_0(z^4+9h(x)^4)\lambda_0 h'(x)}{h(x)^2}\bigg]
	\end{align}
\begin{figure*}
	\includegraphics[width=1\textwidth]{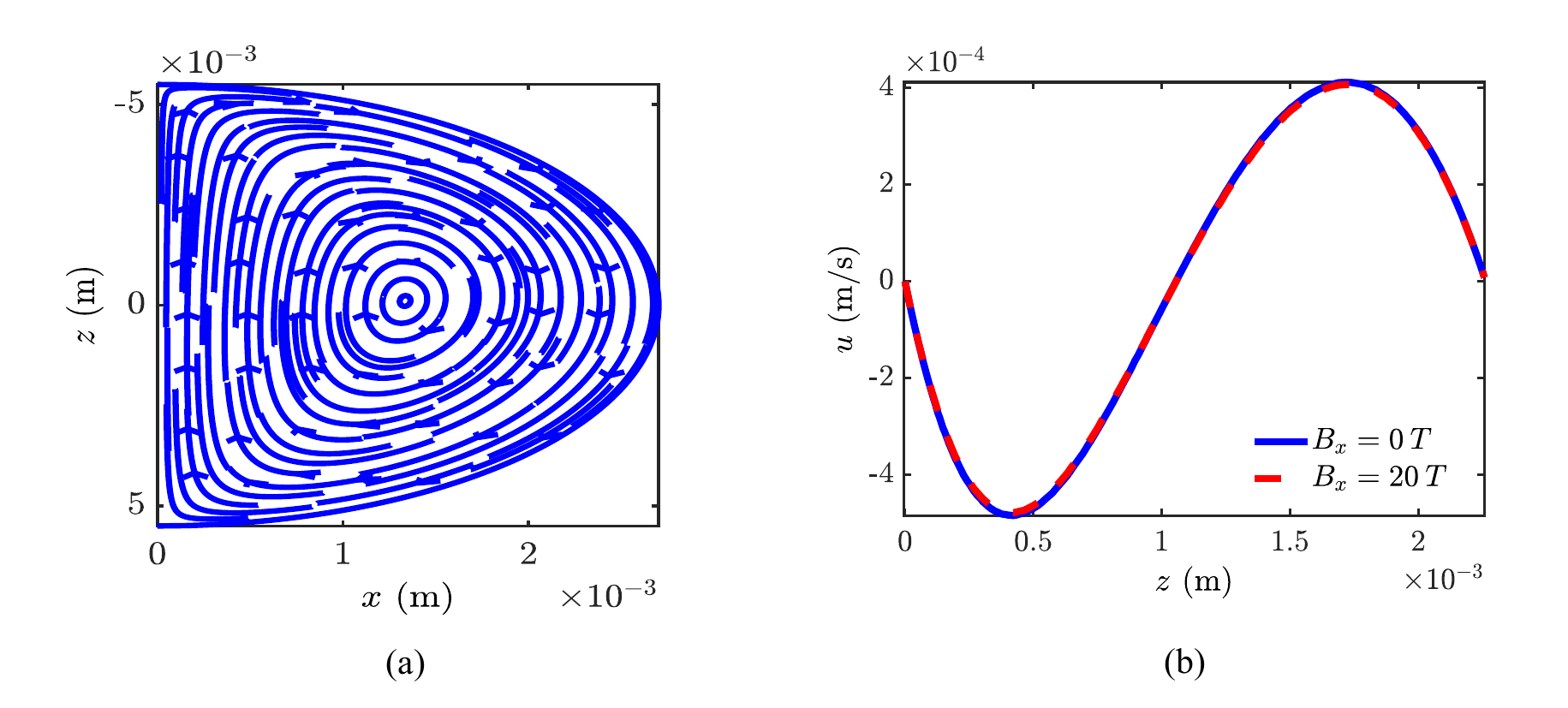}
	\caption{ \label{FIG. 12}(a) Velocity streamlines of AH flow in standing position in the presence of uniform magnetic field $B_x=20$ T (b) Effect of magnetic field on the velocity $(u)$ within the anterior chamber at $x=3$ mm.}
\end{figure*}

	The arbitrary functions $A_1 (x),\hspace{2pt} A_2 (x), \hspace{2pt} A_3 (x)$ and $A_4 (x)$ are obtained using the boundary conditions as done earlier. These are given in the Appendix \ref{Appendix B}.
	We rescale the velocity back to its actual form and obtain
	\begin{equation}\label{eq:55}
		u = \frac{1}{\lambda_0} (\tilde u_0 + \eta \tilde u_1) = u_0 + u_1 \quad \text{and} \quad	w = \frac{1}{\lambda_0} (\tilde w_0 + \eta \tilde w_1) = w_0 + w_1
	\end{equation}

	FIG. \ref{FIG. 7} depicts the stream function of the AH flow for this condition. The warm fluid near the iris rises upwards till it encounters solid corneal surface. As a result, flow in this case is characterized by single clockwise vortex. The symmetry of the flow domain breaks in standing position and thus the flow profile is different from that in the sleeping eye position. The contour plots of $x$-component of velocity obtained numerically and analytically can be found in the Supplementary material (FIG. S4). We see that though $\epsilon$ is not too low, the analytical solution matches well with the numerical solution. The numerical simulations reveal that the maximum magnitude of AH velocity $(u)$, driven solely by buoyancy, reaches $6.61*10^{-4}$  m/s (FIG. S5 (a)). On the other hand, the maximum magnitude in the presence of magnetic field is $5.04*10^{-4}$  m/s (FIG. S4 (a)). Therefore, a uniform magnetic field of 20 T leads to a $23.71 \%$ decrease in maximum AH velocity.

	We compare the analytical and numerical solution quantitatively in the presence of magnetic field in FIG. \ref{FIG. 8}(a) at $x=3$mm. The velocity obtained from both approaches matches closely. To show an improved agreement, the comparison between the analytical and numerical solution has been given in Supplementary material (see FIG. S2 (b)) for small $\epsilon=0.045.$ FIG. \ref{FIG. 8}(b) depicts the effect of magnetic field on AH velocity $(u)$ at $x=3$ mm. Here again as the strength of the uniform magnetic field increases, the opposing body force $\sigma B_z^2 u$ increases leading to lower velocity. 
	
	FIG. \ref{FIG. 9} shows numerically obtained shear stress profile on the corneal and the iris walls in standing position. In the standing case, the iris surface experiences higher shear stress compared to the corneal surface. However, it is not high enough to detach the pigmentary particles from iris. The comparison between numerically and analytically obtained shear stress has been given in the Supplementary material. These match well for small $\epsilon$ value as shown in FIG. S3 (b).

	\subsection{Influence of uniform magnetic field in $x$-direction on AH dynamics }
	\subsubsection*{B.1 Supine position}
	$\left.\begin{array}{l} \boldsymbol{g} = (0,-g), \boldsymbol{B}_{\text{ext}} = (B_x,0) \end{array}\right.$\\
	
	In this case, gravity acts in the negative $z$-direction while the magnetic field acts along the $x$-direction. Eliminating pressure from the governing equations (\ref{eq:9})-(\ref{eq:12}), we obtain
	\begin{equation}\label{eq:56}
		\frac{\partial^4 u}{\partial z^4} - \lambda_2 \frac{\partial^2 u}{\partial x^2} - \frac{\lambda h'(x)}{h(x)^2} = 0
	\end{equation}
	where  $\lambda_2 = \frac{\sigma B_x^2}{\mu}$ and $\lambda = \frac{\rho_0 g \alpha (T_h-T_c)}{\mu}$.

	 FIG. \ref{FIG. 10}(a) shows the numerically obtained streamline plot when a magnetic field of magnitude $20$ T is applied in $x$-direction. The flow is characterised by two counter rotating vortices. In this case, the magnitude of maximum of AH velocity is $11.8*10^{-5}$  m/s as shown in the FIG. S6 (a) in Supplementary material. FIG. \ref{FIG. 3}(b) reveals that maximum magnitude in the absence of magnetic field is $14.1*10^{-5}$  m/s. Therefore, a uniform magnetic field leads to a $16.31 \%$ decrease in AH velocity. FIG. \ref{FIG. 10}(b) depicts the effect of magnetic field on velocity $(u)$ of AH at $x=3$ mm. The Lorentz force in this case is given by $-\sigma B_x^2 w$. The magnetic force is weaker as it depends on $z$-component of velocity $(w)$. Therefore, the effect of the magnetic field applied in $x$-direction is less compared to that applied in the $z$-direction. FIG. \ref{FIG. 11} represents the shear stress on the cornea and the iris wall. The shear stress is too small to detach the pigmentary particles.

	\subsubsection*{B.2 Standing position}
	$\left.\begin{array}{l} \boldsymbol{g} = (g,0), \boldsymbol{B}_{\text{ext}} = (B_x,0) \end{array}\right.$\\
	
	In this case, both gravity and magnetic field act in the $x$-direction. Eliminating pressure from the governing equations (\ref{eq:9})-(\ref{eq:12}), we obtain
	\begin{equation}\label{eq:57}
		\frac{\partial^4 u}{\partial z^4} - \lambda_2 \frac{\partial^2 u}{\partial x^2} = 0
	\end{equation}
	where  $\lambda_2 = \frac{\sigma B_x^2}{\mu}$. In this particular case, the term due to buoyancy does not appear in equation (\ref{eq:57}).
\begin{figure}
	\includegraphics[width=0.45\textwidth]{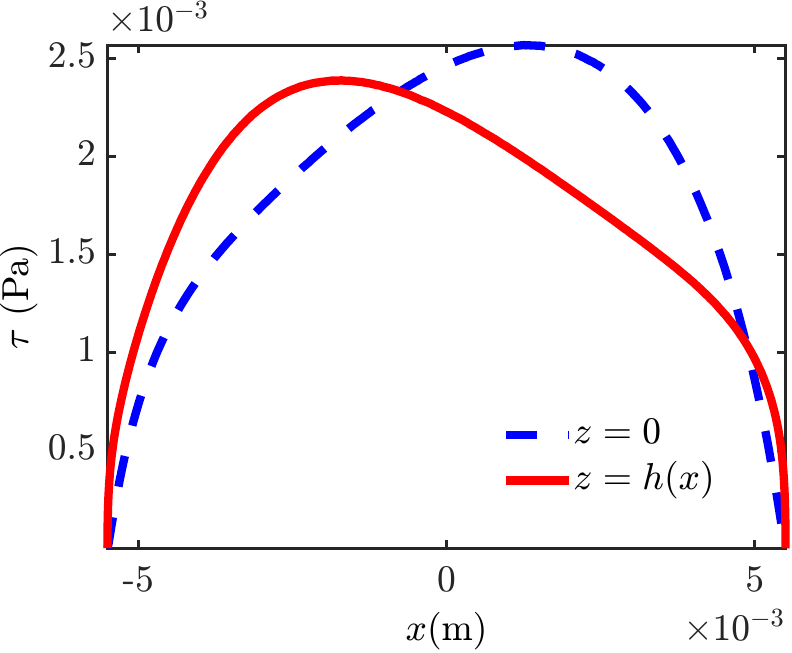}
	\caption{\label{FIG. 13}Shear stress on iris $(z=0)$ and cornea $(z=h(x))$ in the presence of uniform magnetic field $B_x=20$ T in standing position.}
\end{figure}
	
	A magnetic field of magnitude $20$ T is applied in $x$-direction across the eye. FIG. \ref{FIG. 12}(a) shows the numerically obtained streamlines in standing position. In this case, the magnitude of maximum AH velocity is $6.49*10^{-4}$  m/s as shown in FIG. S6 (b) in Supplementary material. In the absence of a magnetic field the maximum velocity is $6.61*10^{-4}$  m/s as shown in FIG. S5 (a) in Supplementary material. Therefore, a uniform magnetic field here leads to only $1.82$ \% decrease in AH velocity. FIG. \ref{FIG. 12}(b) depicts the effect of magnetic field on velocity $(u)$ at $x=3$ mm. The solid blue line represents the variation of $x$-component of AH velocity $(u)$ only due to buoyancy. The red dashed line shows the $x$-component of velocity $(u)$ in the presence of magnetic field. The magnetic field has a negligible effect on AH velocity. FIG. \ref{FIG. 13} represents the shear stress profile on cornea and iris walls. The iris experiences higher shear stress compared to the cornea. Here again the shear stress is insufficient to cause detachment of pigmented particles.
	\section{\label{secIV}Conclusions}
	In this study, we have investigated the dynamics of aqueous humor (AH) in the presence of a uniform magnetic field. While the iris and cornea in the human eye are soft tissues, these can be taken as solid boundaries for the AH flows encountered. This assumption is reasonable as the AH velocity in the anterior chamber (AC) and shear stresses are small. The other important modification we have made is not assumed that the pressure drop is hydrostatic like earlier researchers \citep{canning2002fluid}. We have exploited the fact that pressure is a state variable and the cyclic integral of the change in pressure is zero along a closed path. We use lubrication approximation and determined the flow using regular perturbation method to get an analytical solution. This enables us to obtain solutions solely due to buoyancy and identify the effect of the magnetic field explicitly. The analytical solutions were verified by numerical simulation using COMSOL Multiphysics \textsuperscript \textregistered.\\
	Additionally, we observed that the flow patterns in the eye depend on its orientation. In the standing position, the flow manifests as a single clockwise vortex with its centre in the middle of the anterior chamber, whereas in the supine position, the flow occurs as two counter-rotating vortices symmetric about $x = 0$. We found that the shear stress exerted due to flow of AH has small magnitude and it is insufficient to cause detachment of pigmented particles from the iris and corneal surface. In supine position, corneal wall experiences higher shear stress whereas iris experiences higher shear stress in standing position. The heat generated due to magnetic field is negligible and thus magnetic field should not cause any damage to the eye.

	Although the presence of a magnetic field affects AH flow, the buoyancy forces remain dominant. Therefore, magnetic field does not affect the flow patterns qualitatively. The external magnetic field affected AH velocity significantly when it is oriented in $z$-direction. This is attributed to the magnitude of the Lorentz force, which is higher when the magnetic field is oriented along the z-direction. In the supine position, a magnetic field of strength 20 T in the $z$-direction resulted in a $26.24$ \% decrease in AH velocity. In contrast, the same magnetic field led to a $23.71$ \% decrease in AH velocity in the standing position. When the magnetic field is oriented in the $x$-direction, it is less effective. In the sleeping position, a magnetic field of $20$ T in $x$-direction caused a net decrease of merely $16.31$ \% in AH velocity. In the standing position, the same magnetic field decreased the AH velocity by just $1.82$ \%. This study provides us insights on modifications of AH dynamics under the influence of a magnetic field.

	Understanding the influence of magnetic fields on AH dynamics could lead to the development of new techniques for managing intraocular pressure. This is particularly beneficial for patients with glaucoma. Magnetic field therapy could be fine-tuned to optimize AH flow and reduce pressure. This approach offers a potentially non-invasive treatment option. By strategically orienting magnetic fields, medications can be directed more precisely within the anterior chamber. This precision can improve treatment efficacy for various ocular conditions.
	
	\section*{Supplementary material}
	The supplementary material includes the validation of lubrication approximation, the calculation for pressure integral in standing position and shear stress validation for small AC depth.
	\section*{Acknowledgments}
	The authors thank Indian Institute of Technology, Madras for providing the research facilities. We also thank the Prime Minister’s Research Fellows scholarship for funding this work. We thank Mr. Sambhu Anil for his valuable suggestions during the research.
	\section*{Data availability}
	The data that support the findings of this study are available 	from the corresponding author upon reasonable request.
	
	\appendix 
	\renewcommand\thefigure{\thesection.\arabic{figure}}  
	\setcounter{figure}{0} 
	\section{Calculation for pressure integral}\label{Appendix A}
\begin{figure}[H]
		\includegraphics[width=0.45\textwidth]{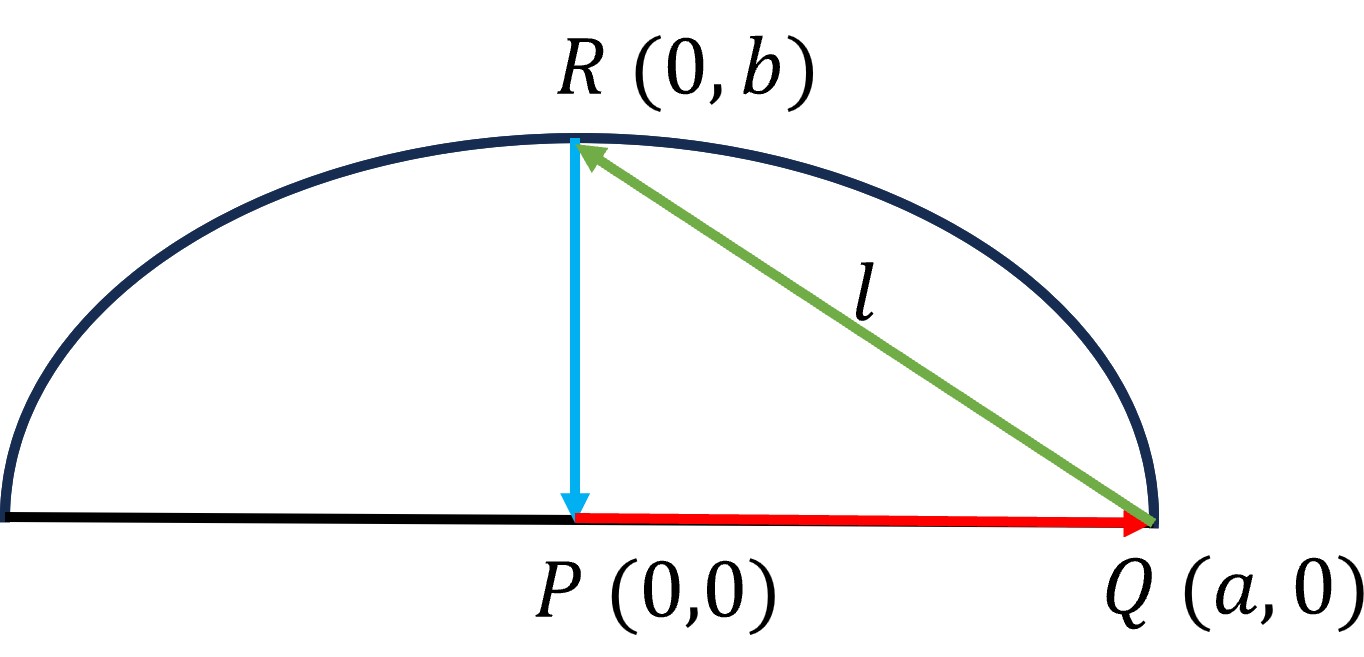}
		\caption{\label{FIG. A.1}The closed contour surface PQRP along which we perform integration to obtain the constant $c_5$.}
		\label{sleep_pressure}
	\end{figure}

	FIG. \ref{FIG. A.1} represents the closed contour PQRP along which we perform integration to obtain the coefficient $c_5$.

	\begin{equation} \label{eq:A1}
		\int_P^Q\frac{\partial p}{\partial x}dx+\int_Q^R \hat{\mathbf{t}}\cdot\nabla p\frac{\partial l}{\partial x}dx+\int_R^P\frac{\partial p}{\partial z}dz=0.
	\end{equation}
	From $x$-momentum balance
	\begin{equation}
		\frac{\partial p}{\partial x} = \mu \frac{\partial^2 u_0}{\partial z^2}.
	\end{equation}
	From $z$-momentum balance
	\begin{equation}
		\frac{\partial p}{\partial z} = -\rho_0 g \left(1 - \alpha (T_h-T_c) \left(1-\frac{z}{h(x)}\right)\right).
	\end{equation}
	The equation of line QR is given as $z= -\frac{b}{a}(x-a)$ and $\frac{\partial l}{\partial x}=\left(1+\left(\frac{\partial z}{\partial x}\right)^2 \right)^{0.5}$. The scalar function defining the line QR is given by
	
	$$	G(x,z) = z + \frac{b}{a}(x-a).$$
	
	The unit normal and tangent vector to the line QR is obtained as
	
	\begin{equation}
		\hat{\mathbf{n}}=\frac{\nabla G}{|\nabla G|} = \frac{1}{\sqrt{1 + (b/a)^2}} \left(\mathbf{\hat{e}}_z + \frac{b}{a} \mathbf{\hat{e}}_x\right) = n_{x} \mathbf{\hat{e}}_x + n_{z} \mathbf{\hat{e}}_z,
	\end{equation}
	
	\begin{equation}
		\hat{\mathbf{t}} = \frac{1}{\sqrt{1 + (b/a)^2}} \left(\frac{b}{a}\mathbf{\hat{e}}_z -  \mathbf{\hat{e}}_x\right) = t_{x} \mathbf{\hat{e}}_x + t_{z} \mathbf{\hat{e}}_z.
	\end{equation}
	
	The term $\hat{\mathbf{t}} \cdot \nabla p  \frac{\partial l}{\partial x}$ is
	
	\begin{equation}
		\hat{\mathbf{t}} \cdot \nabla p  \frac{\partial l}{\partial x} = -\frac{\partial p}{\partial x}+\frac{b}{a}\frac{\partial p}{\partial z}.
	\end{equation}
	Substituting the values of  $\frac{\partial p}{\partial x},\frac{\partial p}{\partial z}$ and $\hat{\mathbf{t}} \cdot \nabla p  \frac{\partial l}{\partial x}$ into equation (\ref{eq:A1}), we obtain
	
	\begin{equation}
		c_5 = \frac{- b\left(\lambda \mu (18-5\pi) + 5g\rho_0 (4-\alpha(-5+\pi)(T_c-T_h))\right)}{ 20\mu\left(\int_{x=0}^{x=a} \frac{dx}{(a^2-x^2)^{3/2}}\right)}=0.
	\end{equation}
	
	\section{Values of arbitrary constants for standing position}\label{Appendix B}
	\begin{equation}
		c_1(x) = \frac{\lambda \lambda_0}{2} + \frac{c_5}{(a^2-x^2)^{3/2}}
	\end{equation}
	\begin{equation}
		c_2(x) = -\frac{\lambda \lambda_0 h(x)}{12} - \frac{c_5 h(x)}{2(a^2 - x^2)^{3/2}}
	\end{equation}
	\begin{equation}
		c_3(x) =0
	\end{equation}
	\begin{equation}
		c_4(x) = 0
	\end{equation}
	\begin{equation}
		c_5 = 0
	\end{equation}
	\begin{equation}
		A_1(x) = \frac{A_5}{h(x)^3} - \frac{1}{120} \lambda_0 h(x) \left( 30c_2(x) + h(x) \left( 9c_1(x) - 2 \lambda \lambda_0 \right) \right)
	\end{equation}
	
	\begin{align}
		A_2(x) = -\frac{1}{120} h(x) \bigg( 120A_1(x) + \lambda_0 h(x)&  \bigg( 20c_2(x) + h(x) &\\ \nonumber \bigg(  5c_1(x) - \lambda \lambda_0 \bigg) \bigg) \bigg)
	\end{align}
	
	\begin{equation}
		A_3(x) =0
	\end{equation}
	\begin{equation}
		A_4(x) =0
	\end{equation}
	\begin{equation}
		A_5(x) =0
	\end{equation}
	
	\section{Justification for solid wall assumption } \label{Appendix C}
	The wall shear stress is proportional to the wall shear rate and is given by equation 
	\begin{equation}
		\tau = \bf{\hat{n}} \cdot \mathbf{S} \cdot \hat{t} 
	\end{equation}
	The scalar function which defines the corneal surface is given by 
	$$F(x,z) = z-h(x)$$
	On evaluating, we obtain
	\begin{equation}\label{eq10}
		\mathbf{\hat{n}} = \frac{-h'(x) \bm{\hat{e}}_x + \bm{\hat{e}}_z}{\sqrt{1 + h'(x)^2}} = n_x \bm{\hat{e}}_x + n_z \bm{\hat{e}}_z
	\end{equation}
	\begin{equation}\label{eq11}
		\mathbf{\hat{t}} = \frac{\bm{\hat{e}}_x  + h'(x) \hat{e}_z}{\sqrt{1 + h'(x)^2}} = t_x \bm{\hat{e}}_x  + t_z \bm{\hat{e}}_z
	\end{equation}
	\begin{equation}\label{eq12}
		\mathbf{S} = S_{xx} \bm{\hat{e}}_x \bm{\hat{e}}_x + S_{xz} \bm{\hat{e}}_x \bm{\hat{e}}_z + S_{zx} \bm{\hat{e}}_z \bm{\hat{e}}_x + S_{zz} \bm{\hat{e}}_z \bm{\hat{e}}_z
	\end{equation}
	where, $S_{xx} = 2\mu \frac{\partial u}{\partial x}$, $S_{xz} = S_{zx} = \mu \left( \frac{\partial u}{\partial z} + \frac{\partial w}{\partial x} \right)$, $S_{zz} = 2\mu \frac{\partial w}{\partial z}$\\
	This gives 
	\begin{equation}
		\tau = n_xS_{xx} t_x+ n_zS_{xz} t_x+ n_xS_{zx} t_z+ n_zS_{zz} t_z.
	\end{equation}
	
	For the corneal surface, the $x$ and $z$ components of normal and tangent vector are given by
	\begin{align}
n_x = \frac{-h'(x)}{\sqrt{1 + h'(x)^2}}, n_z = \frac{1}{\sqrt{1 + h'(x)^2}},& &\\t_x = \frac{1}{\sqrt{1 + h'(x)^2}}, t_z = \frac{h'(x)}{\sqrt{1 + h'(x)^2}}
	\end{align}
		where as $n_x = 0$, $n_z=-1$, $t_x=-1$, $t_z=0$ for iris surface.

The shear modulus $(G)$ and the Poison’s ratio $(\omega)$ of cornea are 72 kPa and 0.49 \citep{ramier2020vivo,asejczyk2011material}. The Young modulus $(\gamma)$ is evaluated using the relation given by \citep{beer1992mechanics}
	\begin{equation} \label{eqc10}
		\gamma = 2G(1+\omega)
	\end{equation}
	This yields a Young’s modulus of the cornea as 0.214 MPa which is consistent with with the experimentally measured value reported by Song et al \cite{song2022measuring}. The maximum shear stress that a material can withstand before failure is known as shear strength $(\tau_{max})$. For the cornea this is obtained as 
	\begin{equation}\label{eqc11}
\tau_{max} = k*G
	\end{equation}
Here, $k$ is a proportionality constant that depends on material. For biological tissues, $k$ is typically taken to be in the range of 0.5 to 0.6. Assuming $k=0.5$, the shear strength $(\tau_{max})$ for cornea is estimated to be 36 kPa. The Young's modulus of normal iris tissue is 0.85 kPa \citep{narayanaswamy2015young}. The Poisson's ratio $\omega$ for iris lies between $0.42-0.47$ \citep{aloy2017estimation}. Taking $\omega=0.45$ and substituting into equations (\ref{eqc10}-\ref{eqc11}), the shear strength for iris is obtained as 0.14 kPa. Hence the shear strength $(\tau_{max})$  for iris is  4.5 kPa. The shear stress experienced by the corneal and iris wall due to aqueous humor circulation are significantly small $O(10^{-4})$.
Therefore, the shear strength for iris and cornea is significantly higher than the shear stress experienced due to AH flow. Hence, the cornea and iris can be considered rigid in the computations.
	
	\section*{References}
	\bibliography{Eye_ref}
\end{document}


\title{\textbf{Supplementary Material}\\ \vspace{ 10 pt} \textbf{Effect of Magnetic Field on Aqueous Humor Flows Inside Anterior Chamber of Human Eye}}
	\author[1]{Deepak Kumar}
	\author[1]{Subramaniam Pushpavanam\thanks{Corresponding author: spush@iitm.ac.in}}
	\affil{ Department of Chemical Engineering, Indian Institute of Technology Madras, Chennai 600036, India}
	\maketitle
	
	\newcommand{\beginsupplement}{%
		\setcounter{table}{0}
		\renewcommand{\thetable}{S\arabic{table}}%
		\setcounter{figure}{0}
		\renewcommand{\thefigure}{\textbf{S\arabic{figure}}}%
		\setcounter{equation}{0}
		\renewcommand{\theequation}{S\arabic{equation}}%
	}
	\renewcommand{\figurename}{FIG.}

	\beginsupplement
		\section*{S.1 Grid independency test: }
\begin{figure*}[h] 
		\centering
		\includegraphics[width=0.9\textwidth]{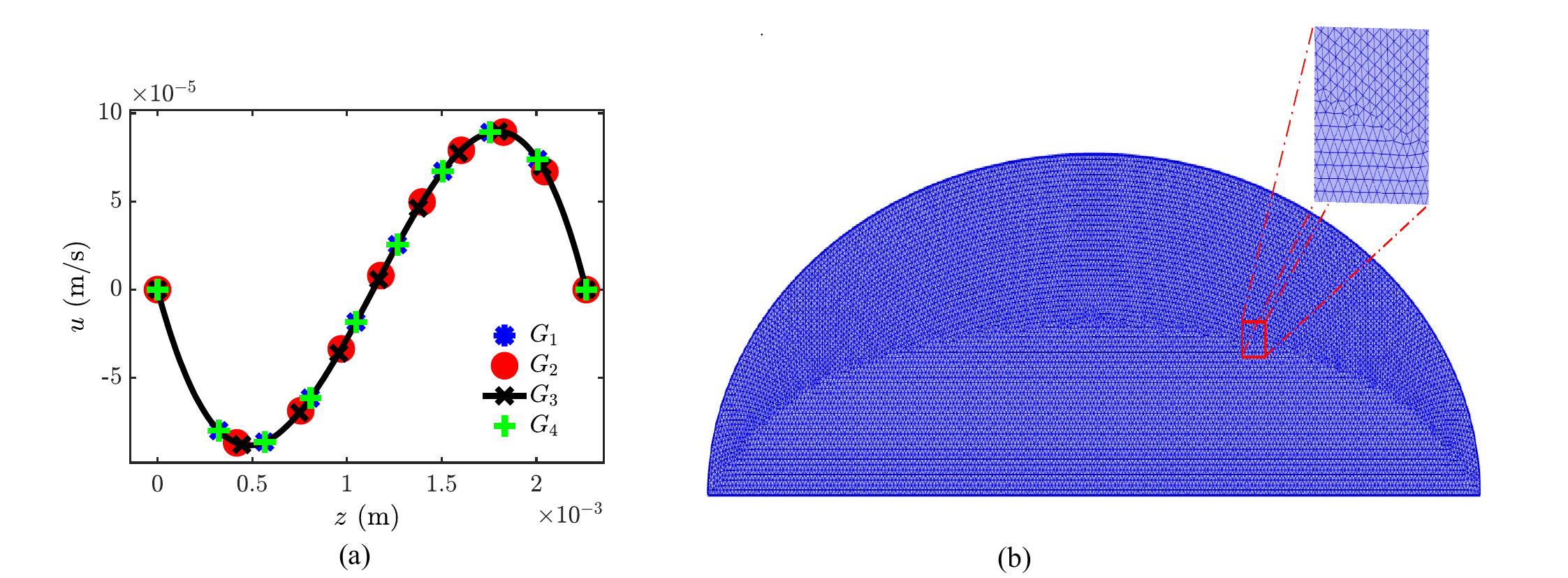}
		\caption{\label{mesh} (a) Effect of mesh on velocity $(u)$ at $x = 3$ mm in supine position of eye under a magnetic field $B_z=20$ T (b) computational mesh $(G_3)$ for simulation in COMSOL Multiphysics 6.2\textsuperscript \textregistered.}
	\end{figure*}

	We have carried out simulations for 5 different grids. The details of these grids have been given in Table \ref{table:grid_data}. The variation of velocity of the AH with $z$ at $x = 3$ mm is calculated for each grid size. The value of  $u_{max}$ remains almost unchanged for grids $G_3, G_4, G_5$ as shown in FIG. \ref{mesh}(a). Hence, we select $G_3$ for computations in our work. Figure \ref{mesh}(b) shows the fine mesh grids  $G_3$ used in the computation. 
	\begin{table}[h!]
		\centering
		\resizebox{\textwidth}{!}{
			\begin{tabular}{|c|c|c|c|c|}
				\hline
				\textbf{Grid} & \textbf{No of domain elements} & \textbf{Minimum element size (m)} & \textbf{Maximum element size (m)} & $\mathbf{u_{\text{max}} \, (m/s)}$ \\
				\hline
				$G_1$ & 1572 & $1.05 \times 10^{-7}$ & $2 \times 10^{-4}$ & $8.92 \times 10^{-5}$ \\
				\hline
				$G_2$ & 6248 & $1.05 \times 10^{-7}$ & $1 \times 10^{-4}$ & $8.92 \times 10^{-5}$ \\
				\hline
				$G_3$ & 24662 & $1.05 \times 10^{-7}$ & $0.5 \times 10^{-4}$ & $8.97 \times 10^{-5}$ \\
				\hline
				$G_4$ & 412110 & $1.05 \times 10^{-7}$ & $0.125 \times 10^{-4}$ & $8.97 \times 10^{-5}$ \\
				\hline
				$G_5$ & 1697648 & $1.05 \times 10^{-7}$ & $0.0625 \times 10^{-4}$ & $8.97 \times 10^{-5}$ \\
				\hline
			\end{tabular}
		}
		\caption{Details of five different mesh elements for grid independency test.}
		\label{table:grid_data}
	\end{table}
	
	\newpage
	\section*{S.2 Validation of lubrication approximation: }
	\begin{figure*}[h]
		\centering
		\includegraphics[width=0.9\textwidth]{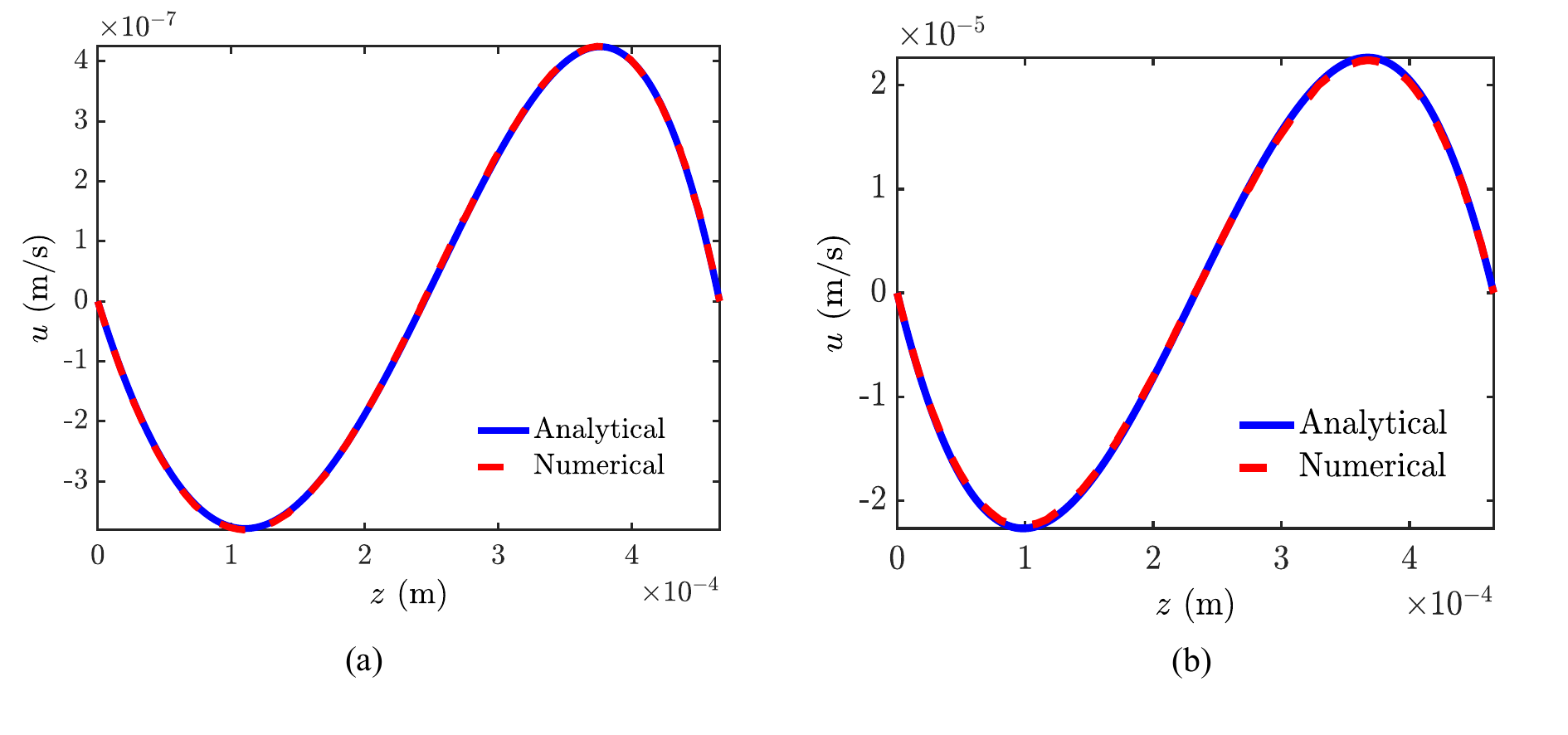}
		\caption{\label{Fig1}Comparison between the analytical and numerical solution of $x$-component of velocity $u$ for $\epsilon=0.0455$ $(L=11$ mm, $b=0.5$ mm) in the presence of uniform magnetic field $B_z=20$ T at $x=0.003$ m (a) supine position (b) standing position.}
	\end{figure*}
	\vspace{1 cm}
		\section*{S.3 Shear stress on walls: }
	\begin{figure*}[h]
		\centering
		\includegraphics[width=0.9\textwidth]{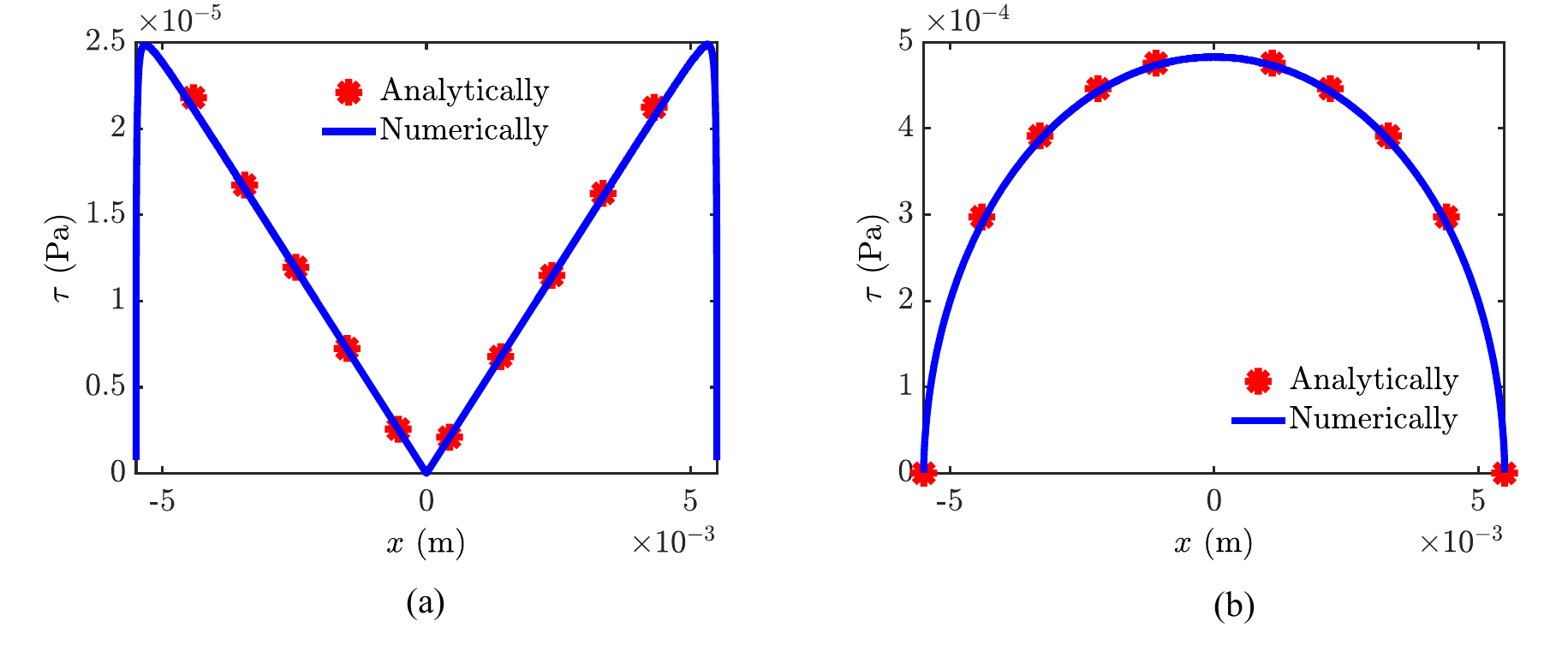}
		\caption{\label{stress} Comparison between the shear stress on the corneal wall obtained using numerical and analytical solution for $\epsilon=0.0455$ $(L=11$ mm, $b=0.5$ mm) (a) supine position (b) standing position.}
	\end{figure*}
	\begin{figure*}[h]
		\centering
		\includegraphics[width=0.9\textwidth]{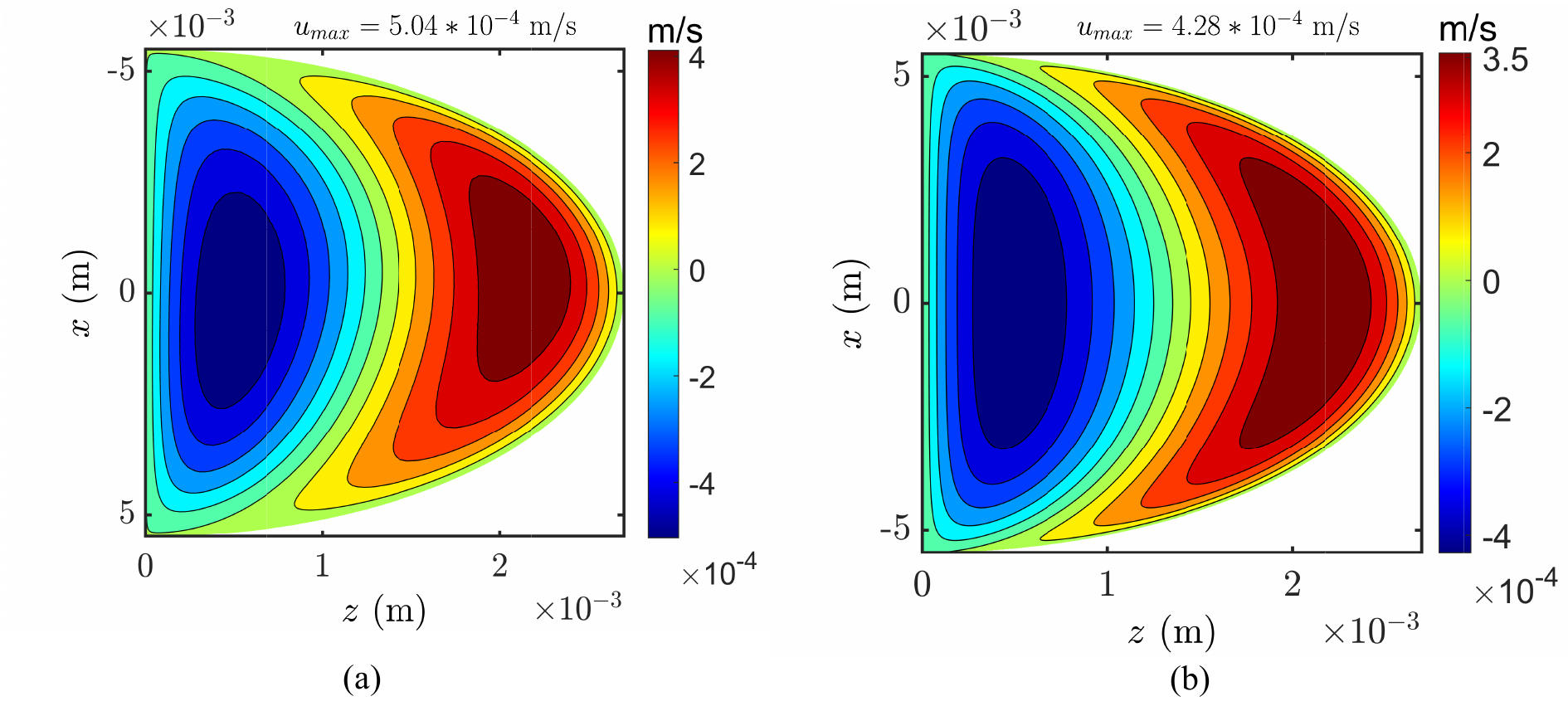}
		\caption{\label{Fig1} Contour plot of $x$-component of AH velocity $(u)$ considering both buoyancy and magnetic field $B_z=20$ T  in standing position obtained (a) numerically (b) analytically.}
	\end{figure*}
		\begin{figure*} 
		\centering
		\includegraphics[width=0.8\textwidth]{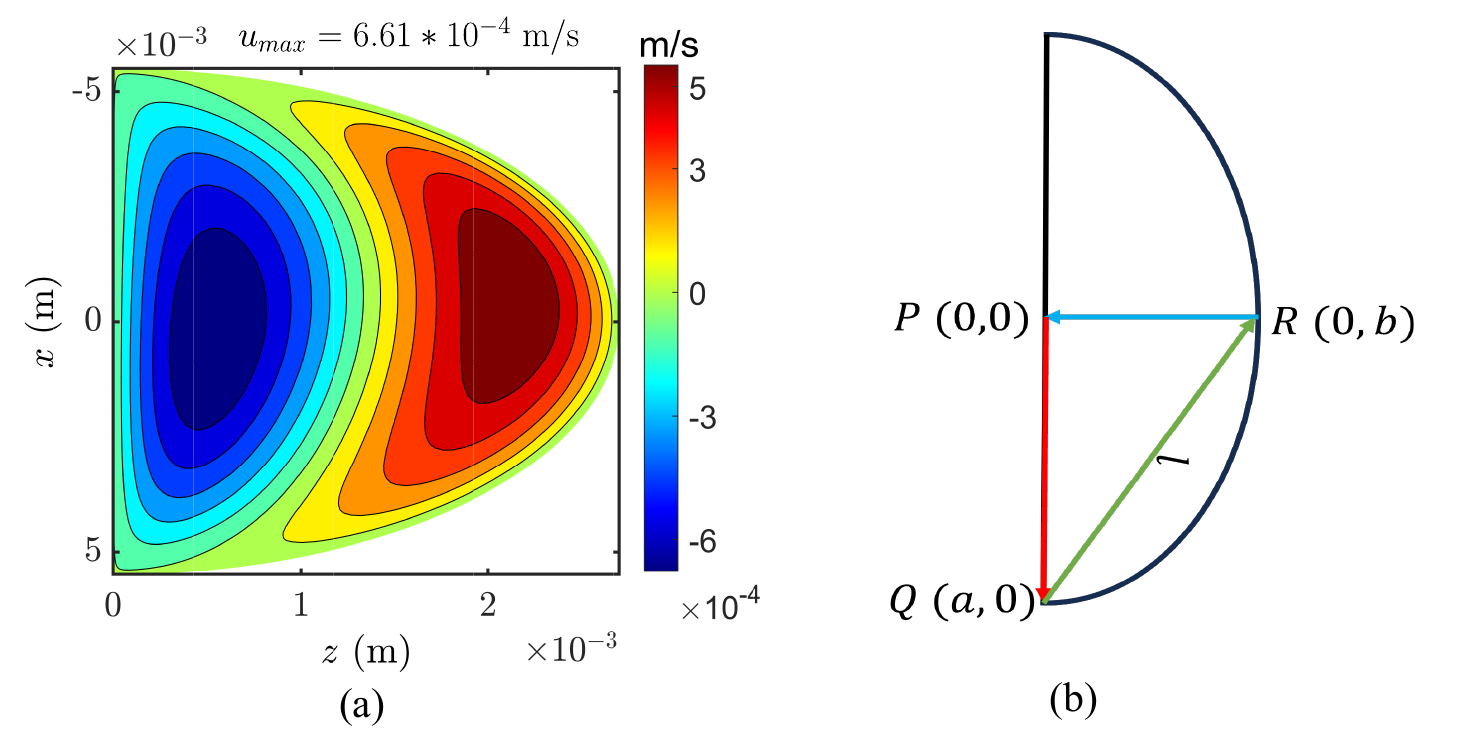}
		\caption{\label{Fig4a}(a) Contour plot of $x$-component of AH velocity $(u)$ in standing position obtained numerically in the absence of magnetic field (b) The closed contour surface PQRP along which we perform integration to obtain the constant $c_5$ in standing position.}
	\end{figure*}

	\newpage
	
	\section*{S.4 Calculation for pressure integral for standing position: }
	FIG. \ref{Fig4a}(b) represents the closed contour PQRP along which we perform integration to obtain the coefficient $c_5$. The equation of line QR is given as $z = -\frac{b}{a}(x-a)$ and $\frac{\partial l}{\partial x}= \sqrt{1+\left(\frac{\partial z}{\partial x}\right)^2}$. The scalar function defining the line QR is given by 
	\begin{equation}
		G(x,z) = z+\frac{b}{a}(x-a).
	\end{equation}
	
	\begin{equation}
		\int_P^Q\frac{\partial p}{\partial x}dx+\int_Q^R \hat{\mathbf{t}}\cdot\nabla p\frac{\partial l}{\partial x}dx+\int_R^P\frac{\partial p}{\partial z}dz=0.
	\end{equation}
	\begin{equation}
		\frac{\partial p}{\partial x} = \mu \frac{\partial^2 u_0}{\partial z^2}+ \rho_0 g \left(1 - \alpha (T_h-T_c) \left(1-\frac{z}{h(x)}\right)\right).
	\end{equation}
	\begin{equation}
		\frac{\partial p}{\partial z} = 0.
	\end{equation}
	
	The unit normal and tangent vector to the line QR is obtained as
	\begin{equation}
		\hat{\mathbf{n}}=\frac{\nabla G}{|\nabla G|} = \frac{1}{\sqrt{1 + (b/a)^2}} \left(\mathbf{\hat{e}}_z + \frac{b}{a} \mathbf{\hat{e}}_x\right) = n_{x} \mathbf{\hat{e}}_x + n_{z} \mathbf{\hat{e}}_z
	\end{equation}
	
	\begin{equation}
		\hat{\mathbf{t}} = \frac{1}{\sqrt{1 + (b/a)^2}} \left(\frac{b}{a}\mathbf{\hat{e}}_z -  \mathbf{\hat{e}}_x\right) = t_{x} \mathbf{\hat{e}}_x + t_{z} \mathbf{\hat{e}}_z
	\end{equation}
	
	The term $\hat{\mathbf{t}} \cdot \nabla p  \frac{\partial l}{\partial x}$ is
	
	\begin{equation}
		\hat{\mathbf{t}} \cdot \nabla p  \frac{\partial l}{\partial x} = -\frac{\partial p}{\partial x}+\frac{b}{a}\frac{\partial p}{\partial z}
	\end{equation}
	Substituting the values of  $\frac{\partial p}{\partial x},\frac{\partial p}{\partial z}$ and $\hat{\mathbf{t}} \cdot \nabla p  \frac{\partial l}{\partial x}$ into equation (A1), we obtain
	
	\begin{equation}
		c_5 = \frac{a(\lambda \mu (-4 + \pi) + g \rho_0 (-4 + \alpha (-6 + \pi)(T_c - T_h)))}{4 \mu \int_{x=0}^{x=a} \frac{1}{(a^2 - x^2)^{3/2}} \, dx}=0
	\end{equation}
		\begin{figure*}[h]
		\centering
		\includegraphics[width=0.8\textwidth]{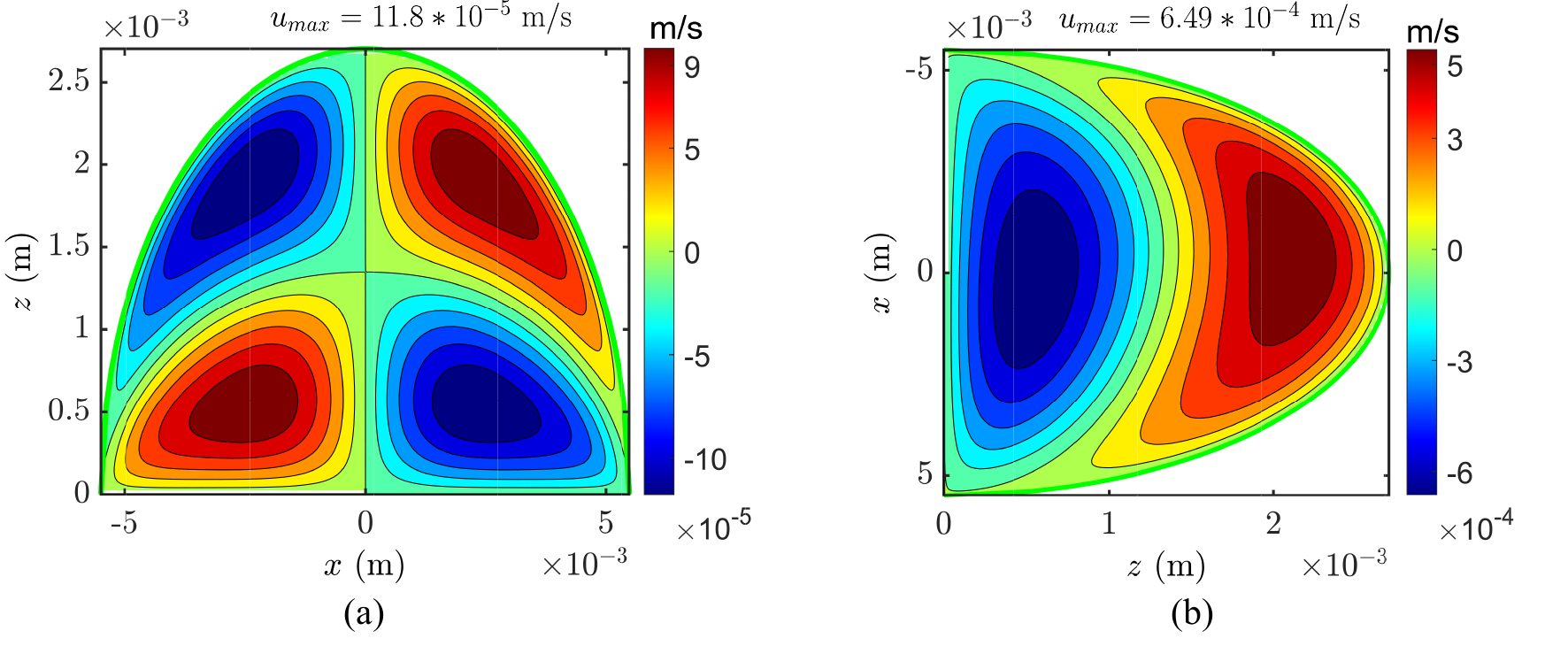}
		\caption{ Contour plot of $x$-component of AH velocity $(u)$ obtained numerically considering both buoyancy and magnetic field $B_x=20$ T (a) supine position (b) standing position.}
	\end{figure*}